\documentclass[conference]{IEEEtran}

\ifCLASSINFOpdf
\else
\fi

\usepackage{amsmath}
\usepackage{amsthm}
\usepackage{xspace}

\newcommand{\x}{\mathbf{x}}

\newcommand{\ours}{$\mathtt{ExpShield}$\xspace}
\newcommand{\ourss}{$\mathtt{ExpShield}$}

\usepackage{subcaption}
\usepackage{listings}
\usepackage[table]{xcolor}
\usepackage{tcolorbox}
\tcbuselibrary{listings}

\definecolor{background}{RGB}{250, 250, 250}
\definecolor{comment}{RGB}{0, 128, 0}
\definecolor{string}{RGB}{208, 16, 64}
\definecolor{keyword}{RGB}{0, 0, 255}
\definecolor{tag}{RGB}{0, 0, 128}
\definecolor{attribute}{RGB}{128, 0, 0}

\lstdefinestyle{HTMLstyle}{
  language=HTML,
  basicstyle=\ttfamily\small,
  backgroundcolor=\color{background},
  keywordstyle=\color{keyword},
  commentstyle=\color{comment},
  stringstyle=\color{string},
  identifierstyle=\color{black},
  morekeywords={style},
  otherkeywords={<,>,/,=},
  frame=none,
  breaklines=true,
  showstringspaces=false,
  tabsize=2
}

\usepackage[utf8]{inputenc} %
\usepackage[T1]{fontenc}    %
\usepackage{hyperref}       %
\usepackage{url}            %
\usepackage{booktabs}       %
\usepackage{amsfonts}       %
\usepackage{nicefrac}       %
\usepackage{microtype}      %
\usepackage{xcolor}         %

\usepackage{lipsum} %
\usepackage{adjustbox} %
\usepackage{cleveref} %
\usepackage{wrapfig} %

\usepackage{pifont}
\usepackage{booktabs}
\usepackage{multirow}
\usepackage[mathlines]{lineno}
\usepackage{enumitem}
\setlist[itemize]{leftmargin=*}
\setlist[enumerate]{leftmargin=*}

\newtheorem{definition}{Definition}

\newtheorem{lemma}{Lemma}
\newtheorem{property}{Property}

\usepackage{tabularx} %
\newcolumntype{Y}{>{\centering\arraybackslash}X} 

\usepackage{tcolorbox}
\tcbuselibrary{skins,breakable}

\newtcolorbox{hypothesisbox}[1][]{
    enhanced, 
    breakable,
    colback=white,          %
    colframe=black!75,      %
    arc=0pt,                %
    boxrule=0.5pt,          %
    left=6pt, right=6pt,    %
    top=6pt, bottom=6pt,    %
    boxsep=0pt,             %
    fontupper=\normalfont,  %
    title={\textbf{Hypothesis}},  %
    #1
}

\usepackage{algorithm}
\usepackage[noend]{algpseudocode}

\usepackage{tikz}
\usepackage{booktabs}

\makeatletter
\newcommand{\linebreakand}{%
  \end{@IEEEauthorhalign}
  \hfill\mbox{}\par
  \mbox{}\hfill\begin{@IEEEauthorhalign}
}
\makeatother

\begin{document}
\title{$\mathtt{ExpShield}$: Safeguarding Web Text from  Unauthorized Crawling and LLM Exploitation}

\author{
    \IEEEauthorblockN{Ruixuan Liu}
    \IEEEauthorblockA{Emory University\\
    ruixuan.liu2@emory.edu}
    \and
    \IEEEauthorblockN{Toan Tran}
    \IEEEauthorblockA{Emory University\\
    viet.toan.tran@emory.edu}
    \and
    \IEEEauthorblockN{Tianhao Wang}
    \IEEEauthorblockA{University of Virginia\\
    tianhao@virginia.edu}
    
    \linebreakand 
    
    \IEEEauthorblockN{Hongsheng Hu}
    \IEEEauthorblockA{Shanghai Jiao Tong University\\
    hongsheng.hu@sjtu.edu.cn}
    \and
    \IEEEauthorblockN{Shuo Wang}
    \IEEEauthorblockA{Shanghai Jiao Tong University\\
    wangshuosj@sjtu.edu.cn}
    \and
    \IEEEauthorblockN{Li Xiong}
    \IEEEauthorblockA{Emory University\\
    lxiong@emory.edu}
}

\IEEEoverridecommandlockouts
\makeatletter\def\@IEEEpubidpullup{6.5\baselineskip}\makeatother
\IEEEpubid{\parbox{\columnwidth}{
		Network and Distributed System Security (NDSS) Symposium 2026\\
		23 - 27 February 2026 , San Diego, CA, USA\\
		ISBN 979-8-9919276-8-0\\  
		https://dx.doi.org/10.14722/ndss.2026.240011\\
		www.ndss-symposium.org
}
\hspace{\columnsep}\makebox[\columnwidth]{}}

\maketitle

\begin{abstract}

As large language models increasingly memorize web-scraped training content, they risk exposing copyrighted or private information.
Existing protections require compliance from crawlers or model developers, fundamentally limiting their effectiveness.
We propose \ours, a proactive self-guard that mitigates memorization while maintaining readability via invisible perturbations, and we formulate it as a constrained optimization problem.
Due to the lack of an individual-level risk metric for natural text, we first propose {\em instance exploitation}, a metric that measures how much training on a specific text increases the chance of guessing that text from a set of candidates—with zero indicating perfect defense. 
Directly solving the problem is infeasible for defenders without sufficient knowledge, thus we develop two effective proxy solutions: single-level optimization and synthetic perturbation.
To enhance the defense, we reveal and verify the memorization trigger hypothesis, which can help to identify key tokens for memorization.
Leveraging this insight, we design targeted perturbations that (i) neutralize inherent trigger tokens to reduce memorization and (ii) introduce artificial trigger tokens to misdirect model memorization.
Experiments validate our defense across attacks, model scales, and tasks in language and vision-to-language modeling. 
Even with privacy backdoor, the Membership Inference Attack (MIA) AUC drops from 0.95 to 0.55 under the defense, and the instance exploitation approaches zero. 
This suggests that compared to the ideal no-misuse scenario, the risk of exposing a text instance remains nearly unchanged despite its inclusion in the training data.

\end{abstract}

\IEEEpeerreviewmaketitle

\section{Introduction}
Building datasets for large language models (LLMs) increasingly depends on crawling and parsing publicly available web content. 
For instance, OpenAI's GPT-3~\cite{brown2020language} was trained on diverse sources including Wikipedia, Common Crawl, books, and articles. 
However, public accessibility does not imply unrestricted usage rights for AI training~\cite{henderson2023foundation, tramer2024position}. 
A critical concern is that LLMs can memorize and reproduce verbatim copyrighted or sensitive content~\cite{carlini2021extracting}, undermining both model generalization and raising serious ethical and legal issues regarding privacy~\cite{carlini2021extracting}, copyright~\cite{freeman2024exploring}, and context collapse~\cite{loh2021social}.

Protecting content from unauthorized use faces two fundamental challenges.
First, preventing web crawling is inherently difficult due to the open nature of the internet.
Sophisticated crawlers can circumvent standard protections by ignoring \texttt{robots.txt} directives, mimicking human browsing patterns, rotating IP addresses, and leveraging distributed networks~\cite{khder2021web}.
Second, existing defenses against verbatim memorization require cooperation from third parties who may not be incentivized to comply.
Data-level protections such as deduplication~\cite{kandpal2022deduplicating} and scrubbing~\cite{lukas2023analyzing} depend on data curators; training-level defenses like differentially private (DP) training~\cite{abadi2016deep} and model alignment~\cite{hendrycks2020aligning} rely on model developers; and inference-level controls such as output filtering~\cite{panaitescu2024can} depend on model curators.\looseness=-1

\begin{figure}[t]
\centering
\includegraphics[trim=23 0 23 0, clip, width=0.9\linewidth]{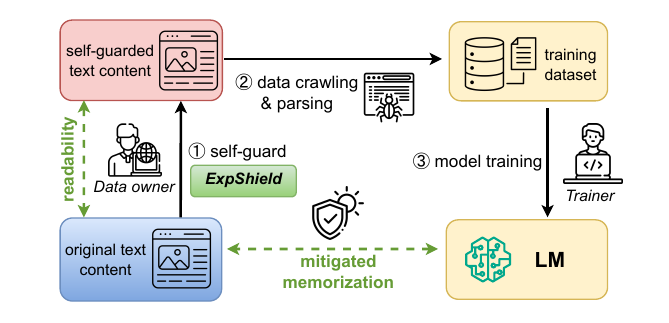}
\caption{Overview of web-text protection. Data owner $\mathcal{O}$ uses \ours to transform original text into protected version before web release. Model trainer $\mathcal{T}$ crawls the protected content for LM training. \ours mitigates verbatim memorization and data leakage in trained LMs.}
\label{fig:overview}
\end{figure}

To address these limitations, we propose a self-guard \ourss~ that empowers data owners with direct control at release time, as shown in \Cref{fig:overview}.
This approach aims to mitigate memorization risk preemptively when the data is misused in LLM training.
Unlike existing countermeasures, our self-guard operates independently without requiring third-party compliance.
Moreover, instead of applying a one-size-fits-all defense, it protects each individual text instance independently.

We formulate the individual-level protection as a constrained bi-level optimization framework with two critical constraints for practical deployment.
The \textit{main objective} is to find a text perturbation that minimizes the adversary's advantage of inferring the protected text when the perturbed version has been used for training.
The \textit{readability constraint} requires that perturbations preserve semantic integrity and rendering consistency for legitimate users—existing privacy-preserving methods~\cite{igamberdiev2023dp,yue2021differential,li2023make} that replace or delete content are unsuitable as they fundamentally compromise text readability.
The \textit{budget constraint} limits perturbation overhead and ensures normal users' experience.

Evaluating and solving the individual-level defense requires a rigorous metric that explicitly quantifies the privacy risk increase caused by model training on protected text instances.
However, enumerating all possible adversaries is impractical, and existing metrics from privacy attacks~\cite{carlini2022membership} and memorization studies~\cite{carliniSecretSharerEvaluating2019} operate at the dataset level or assume uniform risk distributions, failing to account for the inherent variation in memorization susceptibility across natural language.
To address this fundamental limitation, we propose \textit{instance exploitation}—a novel metric that isolates sample-specific memorization from model generalization by calibrating against an informed baseline that has access to all training data except the target instance.

The bottleneck of solving the defense problem is the limited capabilities of data owners, who typically lack access to training algorithms, datasets, or target models, which makes direct bi-level optimization intractable.
Based on previous success in reformulating bi-level problem to single-level~\cite{fowl2021adversarial} or optimization-free solution~\cite{yu2022availability}, we develop two practical proxy solutions: single-level optimization that replaces the target model with open-source proxy models, and optimization-free solution that creates training shortcuts by injecting synthetic perturbation in text.

To instantiate the two solutions effectively, we investigate the fundamental mechanisms underlying memorization in autoregressive language modeling.
Our key insight is the memorization trigger hypothesis: 
specific tokens disproportionately drive the model to rely on memorization over generalization.
We identify memorization triggers as tokens with low prediction confidence under an open-source pre-trained model. 
Intuitively, the low confidence suggests that these tokens are unpredictable given their usual context and stand out as anomalies. 
We verify that such tokens often act as distinctive or rare patterns that the model tends to memorize verbatim if used for training. 

Leveraging this insight, we design targeted perturbations that (i) neutralize inherent memorization triggers through strategic placement of imperceptible elements, and (ii) introduce artificial triggers as adversarial pitfalls that further misdirect the model's memorization on the protected text.
To satisfy the readability and budget constraints, we use invisible Unicode characters and CSS styling to maintain perfect visual fidelity while maximizing defensive efficacy within budget constraints.
We demonstrate our defense mechanism using a fictional webpage example\footnote{Available at: \url{https://github.com/Emory-AIMS/ExpShield-demo}}.

\smallskip
\noindent \textbf{Contributions.} Our contributions are summarized as follows:
\begin{enumerate}[itemsep=1pt]
    \item We formalize individual text protection as constrained bi-level optimization minimizing adversarial advantage while preserving readability. Given data owners' limited capabilities, we develop two practical solutions: synthetic perturbations and single-level optimization with proxy models.\looseness=-1
    \item We propose instance exploitation, a novel privacy metric that quantifies the individual-level memorization risk by calibrating sample-specific exposure against an informed adversarial baseline with access to all other training data, enabling principled design and evaluation of instance-level defenses.
	\item We propose the memorization trigger hypothesis: tokens exhibiting low prediction confidence in general models are the key drivers of memorization. This hypothesis is subsequently validated across diverse language models. Leveraging this insight, we design targeted perturbations that neutralize inherent triggers while introducing artificial triggers as adversarial pitfalls.
    \item Evaluation across various language and vision-language models (124M to 7B parameters) shows that \ours significantly reduces the extraction success rate over $10^5$ attempts and MIA AUC to near-random (0.55) against an informed attack with a privacy backdoor, achieving near-zero instance exploitation with robustness against detection and adaptive scenarios.
\end{enumerate}

\section{Threat Model and Preliminaries}

\subsection{Threat Model}\label{sec:threat}
As shown in \Cref{fig:overview}, we consider two main parties in the web-text protection problem: the data owner $\mathcal{O}$ and trainer $\mathcal{T}$.

\begin{table*}[hbt!]
\centering
\caption{Comparison of trainer assumptions and privacy impact with our defense: 
We target moderate trainers who prioritize data usability without active bypass attempts, as aggressive trainers face prohibitive legal risks and implementation costs for marginal benefits.
(\checkmark: Yes, $\times$: No; N/A: Depends on the bypassing level;
Limited: Our defense is not specifically designed for aggressive trainers, but shows robustness against active bypass in \Cref{sec:robust};
Color coding: \textcolor{green}{Positive}, 
\textcolor{red}{Negative})
}
\label{tab:crawler_comparison}
\resizebox{0.8\textwidth}{!}{
\begin{tabular}{l|cccc|cc|c}
\toprule
\multirow{3}{*}{\textbf{Assumption for $\mathcal{T}$}} & \multicolumn{4}{c|}{\textbf{Crawling Behavior and Cost}} & \multicolumn{2}{c|}{\textbf{w/ ExpShield}} & \multirow{2}{*}{\textbf{Applicable?}} \\ 
\cline{2-7}
& \begin{tabular}[c]{@{}c@{}}Follow\\ Protocol?\end{tabular} & \multicolumn{1}{c|}{\begin{tabular}[c]{@{}c@{}}Active\\ Bypass?\end{tabular}} & \multicolumn{1}{c|}{Cost} & Data Usability & \begin{tabular}[c]{@{}c@{}}Protection\\ Range\end{tabular} & \begin{tabular}[c]{@{}c@{}}Risk\\ Level\end{tabular} & \\
\midrule
\textbf{Conservative} & \checkmark & \multicolumn{1}{c|}{$\times$} & \multicolumn{1}{c|}{\textcolor{green}{No}} & \textcolor{red}{Low} & \textcolor{green}{All} & \textcolor{green}{No Risk} & \textcolor{green}{No hurt} \\ 
\hline
\textbf{Moderate (Our Focus)} & $\times$ & \multicolumn{1}{c|}{$\times$} & \multicolumn{1}{c|}{\textcolor{green}{No}} & \textcolor{green}{High} & \textcolor{green}{Targeted} & \textcolor{green}{Low} & \textcolor{green}{Helpful} \\ 
\hline
\textbf{Aggressive} & $\times$ & \multicolumn{1}{c|}{\checkmark} & \multicolumn{1}{c|}{\textcolor{red}{High}} & \textcolor{green}{High} & N/A & N/A & Limited \\ 
\bottomrule
\end{tabular}
}
\end{table*}

\noindent\textbf{Owner/Defender $\mathcal{O}$ (Data Guarding)}: 
The content owner controls the release of their textual data and seeks to mitigate sample-specific memorization by any LLMs when the released content is subsequently misused for unauthorized training.
The owner cannot foresee potential attacks and has no access to training data or algorithms of potential target model.
The defense targets original content that has not been widely replicated across external sources, ensuring the protection focuses on genuinely unique material.
Crucially, $\mathcal{O}$ does not seek to enhance or degrade overall model performance through the released content. 
This fundamental distinction separates our work from existing unlearnable examples~\cite{fowl2021adversarial,huang2021unlearnable,yu2022availability}, which explicitly aim to impair the model's test performance.

\noindent \textbf{Trainer/Misuser $\mathcal{T}$ (Crawling and Training):} 
This entity systematically crawls web pages to construct training datasets and optimize language model performance. As detailed in \Cref{tab:crawler_comparison}, we model $\mathcal{T}$ as a {\it moderate} actor that disregards standard crawling protocols (e.g., ignoring \texttt{robots.txt} directives) and trains models without implementing privacy-preserving defenses~\cite{abadi2016deep,kandpal2022deduplicating}. Critically, $\mathcal{T}$ does not actively attempt to bypass self-guard mechanisms or deliberately amplify data leakage risks~\cite{bowen2024,liu2024precurious,wen2024privacy}, as such adversarial behavior would incur substantial computational and legal costs. This assumption reflects practical reality, where data misuse typically occurs through negligence rather than malicious intent~\cite{nasr2023scalable}.

While not targeted in this paper, we also describe two other possible $\mathcal{T}$'s. A {\it conservative} $\mathcal{T}$ that adheres to all crawling rules poses no data risk. However, the resulting reduction in training data quantity and diversity compromises usability from the trainer's perspective. Conversely, an {\it aggressive} $\mathcal{T}$ actively bypasses self-guards by detecting and filtering them. Yet, perfectly stripping self-guards without damaging normal text requires significant effort, and the heightened data leakage risk for owners also exposes $\mathcal{T}$ to substantial legal consequences.

\subsection{Background of Language Models}
\noindent\textbf{Model Training.}
Contemporary transformer-based language models~\cite{vaswani2017attention, radford2018improving} employ autoregressive training in both pre-training and fine-tuning phases.
Text from each data owner is tokenized into a sequence $\x = (x_1, x_2, \dots, x_t)$ of length $t$, with the dataset $D = \{\x_1, \dots, \x_n\}$ comprising data from $n$ owners. 
The model's objective is to predict the next token $x_{t+1}$ given the preceding context $(x_1, x_2, \dots, x_t)$.
Training minimizes the negative log-likelihood objective over $T$ tokens:
\begin{align}
\mathcal{L} = -\frac{1}{T}\sum_{t=1}^{T} \log f_\theta(x_{t} | x_{<t}),
\label{eq:llm_loss}
\end{align}
where $f_\theta(x_{t} | x_{<t})$ denotes the conditional probability from model $\theta$'s softmax output, and $x_{<t}$ represents the prefix context.

\noindent\textbf{Content Leakage Risk.} 
In the inference phase, the trained model generates a new text by iteratively sampling $\hat{x}_{t}\sim f_\theta(\cdot | x_{<t})$.
However, previous works show that models can memorize specific training data.
For example, an adversary can efficiently extract training data by querying the target LMs without prior knowledge~\cite{nasr2023scalable}, and the extraction rate increases with more attack attempts~\cite{hayes2024measuring}.
Additionally, membership inference attack (MIA)~\cite{yeomPrivacyRiskMachine2018, carlini2022membership} still stands as a widely used auditing technique with a transparent random data split.
Beyond membership identification, MIA is closely related to data extraction~\cite{carlini2021extracting,nasr2023scalable}.
Thus, we consider and evaluate data leakage risk with both data extraction and MIA.

\section{Individual Text Protection}
\subsection{Problem Definition}\label{sec:invisible}
Given the limited capabilities of the data owner ($\mathcal{O}$ as introduced in \Cref{sec:threat}), the only viable self-guard to mitigate future potential leakage through any model trained on the web text is to embed perturbation in released webpage source code.
More specifically, $\mathcal{O}$ crafts the original web text $\x_i$ with a perturbation $\delta_i$ and releases the guarded text $\x_{\delta_i} = \delta_i \circ \x_i$. 
We formulate the construction of the guarded text as a constrained optimization problem:
\begin{align}
    \min_{\delta_i} \quad & \operatorname{Adv}(\x_i; \theta^*_{\delta_i}, \mathcal{A}) \label{eq:obj} \\
	\text{s.t.} \quad & \theta^*_{\delta_i} \in \arg\min_{\theta} \mathcal{L}(D_{\backslash\x_i} \cup \x_{\delta_i}; \theta), \label{eq:problem_train} \\
    & \text{Multiset}(\x_i) \subseteq \text{Multiset}(\x_{\delta_i}), \label{eq:obj_read} \\
     & \text{EditDist}(\x_i, \x_{\delta_i}) /|\x_i| \leq b, \label{eq:obj_budget}
\end{align}
where $\theta^*_{\delta_i}$ is the  model trained on the released text $\x_{\delta_i}$.

\subsubsection{Main Objective for Defense}
The main goal of individual text protection in \Cref{eq:obj} is to minimize the adversary's advantage $\operatorname{Adv}(\cdot)$ on the protected text $\x_i$ given the trained model $\theta^*_{\delta_i}$ and the attack $\mathcal{A}$.
The attack $\mathcal{A}$ can be membership inference attack, data extraction or other variants of attacks.
And $\operatorname{Adv} (\cdot)$ represents the normalized advantage in the success rate of guessing the secret via attack $\mathcal{A}$ over a baseline guess.
For example, MIA advantage~\cite{yeomPrivacyRiskMachine2018} is defined as $\operatorname{Adv}(x_i)=2\Pr[\hat{b}_i=b_i] - 1$ where $b_i$ is the real membership, $\hat{b}_i$ is predicted by $\mathcal{A}_\text{MIA}$ with a baseline success rate $1/2$.

\subsubsection{Constraint on Perturbation Operation}
The constraint in \Cref{eq:obj_read} ensures that $\x_{\delta_i}$ maintains readability and semantic accuracy for normal web browsers by preserving all text of the original $\x_i$.
Specifically, $\delta_i$ must be an invisible augmentation instead of deleting or replacing characters in $\x_i$.

\subsubsection{Constraint on Perturbation Length}
\Cref{eq:obj_budget} introduces a length constraint that bounds the ratio between the edit distance and the original text length $|\x_i|$ by the perturbation budget $b$.
This constraint limits rendering overhead for normal users.

Our problem formulation for individual text protection differs from previous works.
While unlearnable examples~\cite{huang2021unlearnable,fowl2021adversarial,yu2022availability} use similar bi-level optimization to degrade test performance for image tasks, we target training data leakage reduction for language models and do not seek to degrade test performance.
Another work~\cite{li2023make} extends bi-level minimization~\cite{huang2021unlearnable} for text protection but distorts original text through replacement-based perturbations.
In contrast, we preserve readability via \Cref{eq:obj_read}, which is more challenging but necessary for web content.

\subsection{Evaluating the Individual-Level Defense}\label{sec:metric}

\begin{table}[t]
\centering
\caption{Individual privacy metrics/scores comparison.
}
\label{tab:metric_compare}
\centering
\resizebox{0.9\linewidth}{!}{
\begin{tabular}{l|cccc}
\toprule[1pt]
\multirow{2}{*}{\textbf{Metrics/Scores}} & \begin{tabular}{@{}c@{}}Instance\\Level?\end{tabular} & 
\begin{tabular}{@{}c@{}}Standardized?\end{tabular} & 
\begin{tabular}{@{}c@{}}Calibrated?\end{tabular} &
\begin{tabular}{@{}c@{}}Natural\\instance?\end{tabular} \\
\midrule[0.5pt]
MIA-Loss~\cite{yeomPrivacyRiskMachine2018}   & \checkmark                & $\times$          & $\times$                & \checkmark           \\
MIA-LiRA~\cite{carlini2022membership}        & \checkmark          & $\times$          & \checkmark               & \checkmark           \\
\midrule
Canary  Exposure~\cite{carliniSecretSharerEvaluating2019}           & \checkmark      & \checkmark         & $\times$               & $\times$            \\
\midrule
Instance Exploitation & \checkmark & \checkmark         & \checkmark               & \checkmark   \\
\toprule[1pt]
\end{tabular}
}
\end{table}

Given the defined problem, we need an effective individual-level metric to evaluate how well a solution $\delta_i$ reduces the risk of the protected $\x_i$ as formulated in \Cref{eq:obj}.
Thus, dataset-level metrics such as success rate or TPR~\cite{carlini2022membership} for MIA or extractable rate~\cite{carlini2021extracting} for data extraction are inapplicable.
Besides, it should be:
a) \textbf{standardized} to generally compare risks among different architectures;
b) \textbf{calibrated} to accurately capture the risk improvement caused by model training;
and c) \textbf{efficient} to compute for large language models.

\subsubsection{Standardizing Individual Risk via Log-Rank}
Evaluating \Cref{eq:obj} by considering all attacks is impractical, thus we need a proxy metric for various $\mathcal{A}$.
Log-perplexity is a natural choice~\cite{carliniSecretSharerEvaluating2019} as it represents negative log-likelihood of generating $\x$ given $\theta$.
A small value indicates high extraction probability and easier MIA identification of $\x$.
For equal-length text, loss (defined in \Cref{eq:llm_loss}) and log-perplexity differ only by the factor the sequence length, so we use them interchangeably.
Since loss is not standardized across models, we use exposure~\cite{carliniSecretSharerEvaluating2019} to standardize $\mathcal{L}(\x; \theta)$ by ranking against candidate losses from the same distribution, as in \Cref{def:exposure}.

\begin{definition}[Exposure~\cite{carliniSecretSharerEvaluating2019}]\label{def:exposure}
Given the target model $\theta$, let $\mathbf{rank}_\theta(\x)=|\{\x^\prime \in \mathcal{D}: \mathcal{L}(\x^\prime; \theta) \le \mathcal{L}(\x; \theta) \}|$ represent the rank of $\mathcal{L}(\x; \theta)$ among losses of all samples in the domain $\mathcal{D}$.
The exposure $\mathbf{E}_\theta(\x)$ is defined as 
\begin{align}
\mathbf{E}_\theta(\x) &:= \ln |\mathcal{D}| -\ln \mathbf{rank}_\theta(\x) \\
&= -\ln \frac{\mathbf{rank}_\theta(\x)}{|\mathcal{D}|} \\
&= -\ln \mathbf{Pr}_{\x^\prime \in \mathcal{D}}\left[ \mathcal{L}_\theta(\x^\prime) \le \mathcal{L}_\theta(\x) \right].
\end{align}
\end{definition}

Essentially, \Cref{def:exposure} quantifies the advantage of a model-informed adversary over a baseline adversary in a guessing game.
The baseline attack $\mathcal{A}_\text{base}^\text{unif}$ assumes uniform distribution over all candidates in $\mathcal{D}$ and requires on average $|\mathcal{D}|/2$ guesses to find $\x$.
In contrast, the model-informed attack $\mathcal{A}_\text{target}$ leverages $\theta$ to prioritize candidates with lowest loss, requiring only $\textbf{rank}_\theta(\x)$ guesses on average.
The exposure measures how much the knowledge of $\theta$ reduces the expected effort required for guessing a target secret $\x\in\mathcal{D}$.

\subsubsection{Calibration with Informed Adversary}
The exposure metric was originally designed for fixed-template random canaries, which follow uniform distribution.
However, text has a non-uniform distribution, making the uniform baseline $\mathcal{A}_\text{base}^\text{unif}$ weak and leading to overestimated privacy risks.
For example, commonly occurring text (e.g., phrases partially seen during pre-training) will artificially inflate privacy risk scores.

To address this limitation, we propose a much stronger, informed baseline $\mathcal{A}_\text{base}^\text{info}$ inspired by the worst-case assumptions in differential privacy~\cite{dworkDifferentialPrivacyPractice2019}.
This baseline adversary knows all other training data except $\x$ and can train a reference model $\theta_{\backslash \x} \gets \mathcal{T}(D_{\backslash \x})$ using the same training procedure $\mathcal{T}$.
Since $\theta_{\backslash \x}$ and $\theta$ share identical training procedures (including initialization), they exhibit similar loss distributions.
Thus, one optimal strategy for the informed baseline is prioritizing candidates with the lowest loss according to $\theta_{\backslash \x}$.
We define instance exploitation by calibrating the target model's exposure against this informed baseline, as shown in \Cref{def:exploit}.

\begin{definition}[Instance Exploitation] \label{def:exploit}
Given two datasets $D$ and $D_{\backslash \x}$ and models trained by $\mathcal{T}$ over the two datasets $\theta$ and $\theta_{\backslash \x}$, 
the instance exploitation $\mathbf{Ex}$ is defined as
\begin{align}
\mathbf{Ex}(\x; D, \mathcal{T}) &:= \mathbf{E}_\theta(\x) - \mathbf{E}_{\theta_{\backslash \x}}(\x) \\
&= \ln \frac{\mathbf{Pr}_{\x^\prime \in \mathcal{D}}\left[ \mathcal{L}_{\theta_{\backslash \x}}(\x^\prime) \le \mathcal{L}_{\theta_{\backslash \x}}(\x) \right]}{\mathbf{Pr}_{\x^\prime \in \mathcal{D}}\left[ \mathcal{L}_\theta(\x^\prime) \le \mathcal{L}_\theta(\x) \right]}. \label{eq:exploit_frac}
\end{align}
\end{definition}
Essentially, \Cref{def:exploit} measures the guessing advantage: it quantifies how much easier it becomes to identify $\x$ when the model is trained on it, compared to an informed baseline that knows all other training data.
Mathematically, this corresponds to the ratio $\textbf{rank}_{\theta_{\backslash \x}}(\x)/\textbf{rank}_\theta(\x)$ between the expected number of guesses required by $\mathcal{A}_\text{base}^\text{info}$ and $\mathcal{A}_\text{target}$.

A higher instance exploitation value indicates greater adversarial advantage from training on $\x$.
We define a perfect defense as achieving zero or negative instance exploitation, as formalized in \Cref{pro:perfect}.

\begin{property}[Perfect Defense]\label{pro:perfect}
A defense mechanism is \emph{perfect} with respect to a training algorithm $\mathcal{T}$ and domain $\mathcal{D}$ if, for any instance $\x$, the following holds:
\[
\mathbf{Ex}(\x; D, \mathcal{T}) \le 0 \quad \text{or, equivalently} \quad \mathbf{E}_\theta(\x) \le \mathbf{E}_{\theta_{\backslash \x}}(\x).
\]
\end{property}

\begin{algorithm}[t]
\caption{\textsc{Privacy Game for Informed Inference}}
\label{alg:informed-adv}
\begin{algorithmic}[1]
\Procedure{\textsc{Informed-Inference}}{$\mathcal{T}, D_{\backslash 
x}, \mathcal{D}, \x$}
    \State $\tilde{x} \gets$ \ours$(\x)$ if self-guard; else $\tilde{\x}\gets\x$
    \State $\theta \gets \mathcal{T}(D_{\backslash \x} \cup \{\tilde{\x}\})$ \textcolor{gray}{\textit{// Trainer trains target model}}
    \State $\theta_{\backslash \x} \gets \mathcal{T}(D_{\backslash \x})$ \textcolor{gray}{\textit{// $\mathcal{A}_\text{target}^\text{info}$ trains reference model}}
    \State $\mathcal{A}_\text{target}^\text{info}$  sorts descending $\x_i\in \mathcal{D}$ with $\textbf{Ex}(\x_i; \theta, \theta_{\backslash\x})$
    \State $\hat{\x}\gets\mathcal{A}_\text{target}^\text{info}(D_{\backslash\x}, \mathcal{T}, \mathcal{D})$ 
    \textcolor{gray}{\textit{// Guess from top candidates}}
    \State \Return $\mathcal{A}_\text{target}^\text{info}$ wins if $\hat{\x}=\x$; otherwise fails
\EndProcedure
\end{algorithmic}
\end{algorithm}

\noindent\textbf{Informed Attacks and Reducibility.}
The informed adversary assumptions in our instance exploitation metric naturally lead to stronger attack strategies.
By leveraging the same knowledge (access to $D_{\backslash \x}$ and $\mathcal{T}$), an adversary can construct an enhanced attack $\mathcal{A}_\text{target}^\text{info}$ that prioritizes guessing on candidates with top instance exploitation scores rather than raw loss rankings, as formalized in \Cref{alg:informed-adv}.

This provides theoretical justification for our metric through privacy game reducibility~\cite{balle2022reconstructing, salem2023sok}.
When privacy game $G_1$ is reducible to $G_2$ (i.e., $G_1$ is at most as hard as $G_2$), any defense effective against $G_1$ also protects against $G_2$.
Our exploitation metric captures the advantage of a highly informed adversary $\mathcal{A}_\text{target}^\text{info}$, which is similar to informed MIA attacks reducible to data extraction and standard MIAs.
Thus, defenses that reduce exploitation scores provide protection against a broad spectrum of weaker attacks.
This is particularly relevant in practice, as real-world adversaries rarely access the complete training dataset $D_{\backslash\x}$ and exact training procedure $\mathcal{T}$ assumed by the adversary in our metric.

\noindent\textbf{Connection to DP.}
If a model is trained with differential privacy (DP), by the data-processing inequality, the instance exploitation of each training sample is bounded by its DP budget as shown in \Cref{lemma:dp}. 
The transition from DP to instance exploitation is one-directional because \Cref{def:exploit} is an evaluation metric rather than an algorithm that provides theoretical DP guarantee.
\begin{lemma}\label{lemma:dp}
If $\mathcal{T}$ performs $(\epsilon, \delta)$-DP training, the instance exploitation for any sample $\x$ in any $D$ satisfies $\mathbf{Ex}(\x; D, \mathcal{T}) \leq \epsilon$ with a failure probability of $\delta$.
\end{lemma}

\subsubsection{Approximation for Efficient Estimation}
Now we discuss how to efficiently compute the proposed metric \Cref{def:exploit}.

Computing the exposure $\textbf{E}_\theta(\x)$ exactly requires computing losses for all samples in the domain $\mathcal{D}$, which is inefficient when $|\mathcal{D}|$ is very large.
Given auxiliary data $D_\text{aux} \in \mathcal{D}$ not trained on $\theta$, the loss distribution can be modeled as a skew-normal distribution~\cite{carliniSecretSharerEvaluating2019} with mean $\mu$, standard deviation $\sigma$, and skew $\alpha$.
The exposure can then be efficiently estimated as: 
\begin{align}
\mathbf{E}_\theta(\x) \approx \hat{\mathbf{E}}_\theta(\x) = -\ln \int_{0}^{\mathcal{L}_\theta(\x)} \rho(x)dx, \label{eq:modeling}
\end{align}
where $\rho(x)$ is the continuous density function.

\begin{figure}
    \centering
    \includegraphics[width=0.6\linewidth]{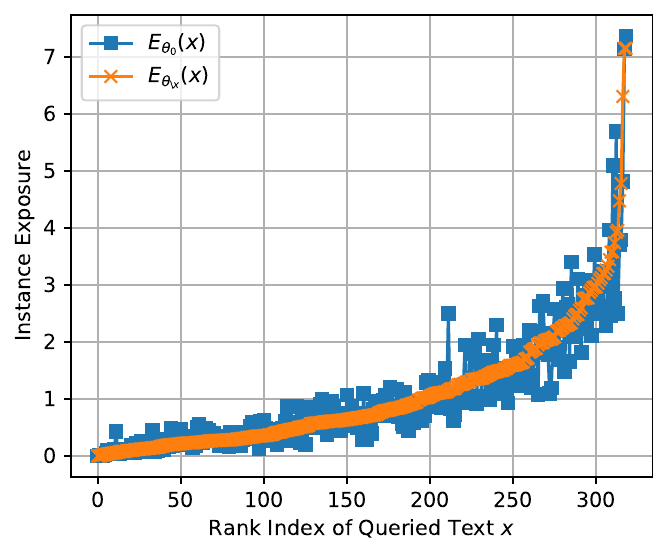}  %
    \caption{Approximation on instance exploitation for efficient estimation. We take the pre-trained GPT-2 as $\theta_\text{pre}$ and fine-tune it with Patient dataset $D_{\backslash \x}$ which excludes the evaluated sample $\x$ for obtaining $\theta_{\backslash \x}$. 
    The modeled skew-normal distribution matches the sampled log-perplexity perfectly because the Kolmogorov-Smirnov goodness-of-fit test~\cite{massey1951kolmogorov} fails to reject the null hypothesis with $p\gg 0.1$.}
    \label{fig:rank_ex}
\end{figure}

Additionally, computing exploitation requires training $\theta_{\backslash \x}$ for each target secret $\x$, which is inefficient for large models.
For protection set $D_\text{pro} \subset D$, we approximate $\theta_{\backslash \x}$ by training on the remaining unprotected data:
\begin{align}
\mathbf{Ex}(\x; D, \mathcal{T}) \approx \hat{\mathbf{Ex}}(\x; D, \mathcal{T}) = \mathbf{E}_\theta(\x) - \mathbf{E}_{\theta_{\backslash D_\text{pro}}}(\x). \label{eq:exp_approx}
\end{align}
When all secrets in $D$ are protected, this equals using the initial model $\theta_\text{pre}$ in place of $\theta_{\backslash \x}$.
As shown in \Cref{fig:rank_ex}, $\mathbf{E}_{\theta_\text{pre}}(\x)$ closely approximates the exact calibration $\mathbf{E}_{\theta_{\backslash \x}}(\x)$ with minimal fluctuation from training randomness and cross-sample influence~\cite{carlini2022privacy}.
The approximation accuracy improves with smaller protection ratios $|D_\text{pro}|/|D|$ due to reduced inter-instance influence.

\subsection{Challenges and Proxy Solution Overview}
We now discuss how to solve the defined problem as the defender $\mathcal{O}$.
Ideally, $\mathcal{O}$ should optimize the bi-level problem in \Cref{eq:obj} for each possible adversary $\mathcal{A}$, which is impractical.
Similarly, due to lack of capability, other alternative proxy objectives are hard to solve as summarized in \Cref{tab:proxy}.

\begin{table}[tb]
\centering
\caption{Summary of Proxy Solutions.}\label{tab:proxy}
\resizebox{0.98\linewidth}{!}{
\begin{tabular}{c|c|c|c}
\toprule[0.9pt]
\textbf{Main Objective} & \textbf{Constraint} & \textbf{Requirement} & \textbf{Defender's Capability} \\
\hline
$\min\operatorname{Adv}(\x; \theta_{\delta}^*, \mathcal{A})$          &  Eq.(2)-(5)   &      $D_{\backslash \x}, \mathcal{T}, \mathcal{A}$            &      $\times$      \\
$\min \mathbf{Ex}(\x)$  & Eq.(2)-(5) & $D_{\backslash \x}, \mathcal{T}, D_\text{aux} $                    &   $\times$    \\
$\max \mathcal{L}_{\theta_{\delta}^*}(\x)$ & Eq.(2)-(5) & $D_{\backslash \x}, \mathcal{T}$ & $\times$ \\
\hline
$\max \mathcal{L}_{\theta_\text{proxy}}(\x_\delta)$ & Eq.(3)-(5) & $\theta_\text{proxy}$ & \checkmark (Our TP-OP) \\
Optimization-free & Eq.(3)-(5) & N/A; $\theta_\text{proxy}$ is optional & \checkmark (Synthetic perturbation) \\
\bottomrule[0.8pt]
\end{tabular}
}
\end{table}

Given the key bottleneck of lacking access to $\mathcal{T}$ and $D_{\backslash\x}$, we propose two practical alternatives:

\noindent\textbf{a) Single-Level Optimization}:
While $\theta_{\delta}^*$ in \Cref{eq:problem_train} is unpredictable, it operates on natural text and shares foundational knowledge with existing open-source models.
We replace $\theta_{\delta}^*$ with an accessible proxy $\theta_\text{proxy}$ and optimize $\max \mathcal{L}_{\theta_\text{proxy}}(\x_\delta)$ given the absence of $\mathcal{A}$ or $D_\text{aux}$.
The intuition is that the perturbations with a high loss on the proxy model are likely to be abnormal patterns against general text and thus can encourage the target model to fit the shortcut.
When a perturbation maximizes the loss on a proxy model, it is likely to mislead the target model.
This single-level approach follows successful precedent in adversarial examples, where bypassing bi-level optimization produces effective poisons~\cite{fowl2021adversarial}.

\noindent\textbf{b) Synthetic Perturbation}:
Discrete optimization over vocabulary incurs substantial computational costs, particularly for long sequences $\x_\delta$ and large models $\theta_\text{proxy}$.
We propose a lightweight alternative using synthetic perturbations that naturally create training shortcuts, encouraging the model to fit $\delta$ rather than memorize $\x$.
This approach builds on recent work demonstrating that synthetic patches can effectively replace bi-level optimized perturbations~\cite{yu2022availability}.
While lacking explicit optimization objectives, we verify in \Cref{sec:verify} that synthetic perturbations implicitly encourage $\max \mathcal{L}_{\theta_\delta^*}(\x)$.

We primarily employ synthetic perturbations for their efficiency and effectiveness, reserving single-level optimization (TP-OP) for scenarios with constrained perturbation budgets.

\section{Self-Guard against Exploitation}

\subsection{Invisibility Strategies}\label{sec:inv}
For \Cref{eq:obj_read}, we consider two strategies to hide perturbation in web page rendering: 1) \textit{invisible style}, including adjusting CSS properties like font size or absolute position for inserting random text; 2) \textit{invisible character}\footnote{\url{https://invisible-characters.com}}, including zero-width and invisible whitespace characters~\cite{boucher2022bad}.
Both invisible characters and styles have been leveraged in attacks~\cite{boucher2022bad, liao2024eia}, while we use for defense purposes.
We elaborate details with a simplified demonstration in \Cref{APP:sec:invisible}.

\noindent \textbf{Robustness against Normal Pre-processing.} 
Crawled content includes visible text, HTML tags, and formatting markers. Standard parsing tools like Beautiful Soup~\cite{richardson2007beautiful} decode entities and strip markup while preserving hidden DOM elements. 
We provide a demo~\cite{expshield-demo-2025} showing that \ours successfully embeds tokens that remain intact in text extracted by four popular web-scraping tools without changing page appearance.

Our defense is robust against other normal pre-processing by its design.
For example, we avoid repeated patterns, so deduplication~\cite{broder1997resemblance}  poses no threat.
And quality filtering~\cite{brown2020language} may trigger removal on the whole sentence, which  enhances protection by preventing training entirely.

\noindent \textbf{Robustness of Active Bypass.}
An adversarial $\mathcal{T}$ may attempt to bypass \ours by perfectly stripping self-guards while preserving original content.
Assuming constant-time $O(1)$ verification per token, the time complexity is $O(T)$, and the tokenized sequence length $T$ of all concatenated text can reach hundreds of billions~\cite{brown2020language}.
Stripping invisible styles incurs larger constant overhead as they permit arbitrary vocabulary insertions, requiring additional operations to recover tokenization boundaries.
While removing invisible characters (e.g., zero-width spaces) has less verification overhead, it maintains $O(T)$ complexity.
A thorough character stripping is possible but it may degrade model robustness~\cite{longpre2024pretrainer,anonymous2025do}.

Advanced bypass strategies involve embedding or perplexity analyses before manual perturbation removal.
We demonstrate robustness against such detection-based attacks in \Cref{sec:robust}.

\subsection{A Basic Random Perturbation for Defense}\label{sec:guarding_via_random}
\begin{table}[t!]
\centering
\caption{Summary of ExpShield variants. 
{OOV} is out-of-vocabulary; {pitfall} means artificially created outlier tokens.
}\label{tab:ours}
\resizebox{0.48\textwidth}{!}{
\begin{tabular}{lcccc}
\toprule
\textbf{Methods} & \textbf{Perturb Location} & \textbf{Filling Strategy} & \textbf{Invisibility} \\
\midrule
UDP (\textsection~\ref{sec:uniform})     & Deterministic & Uniform & Style \\
UNP (\textsection~\ref{sec:uniform})     & Non-Deterministic & Uniform & Style \\
\hline
TP (\textsection~\ref{sec:tp})     & Mem. Trigger & Uniform & Style \\
TP-P (\textsection~\ref{sec:pitfall})     & Mem. Trigger & Outlier pitfall & Style \\
TP-OP (\textsection~\ref{sec:pitfall_opt})    & Mem. Trigger & Optimized pitfall & Style \\
TP-OOV (\textsection~\ref{sec:pitfall_oov})   & Mem. Trigger & OOV pitfall & Character \\
\bottomrule
\end{tabular}
}
\end{table}

\begin{figure}[t!]
    \centering
    \begin{subfigure}[b]{0.245\textwidth}
        \includegraphics[width=\linewidth, trim=0 0 0 0, clip]{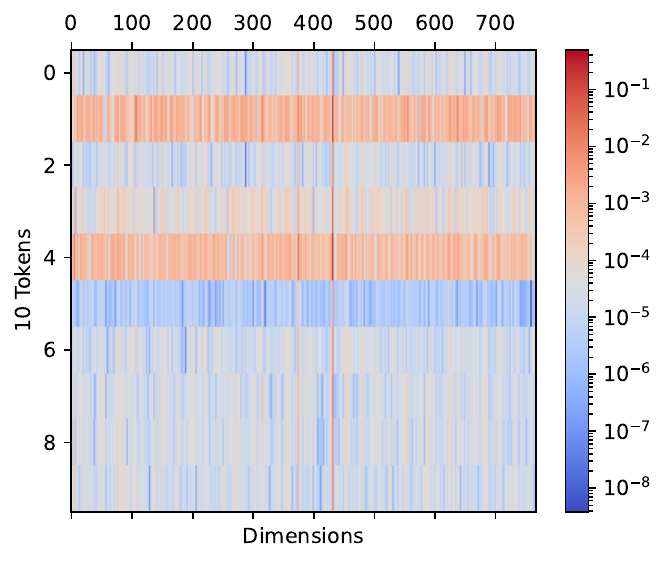}
    \end{subfigure}
    \hfill
    \begin{subfigure}[b]{0.23\textwidth}
        \includegraphics[width=\linewidth, trim=20 0 0 0, clip]{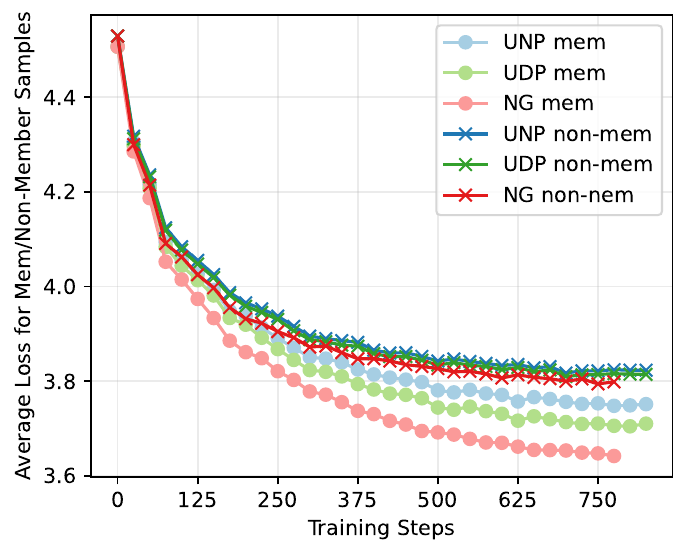}
    \end{subfigure}
    \caption{
    (Left) Difference between embedding gradients of UDP and baseline without perturbation, with x-axis as the embedding dimension and y-axis as the protected token's sequence.
    (Right) Loss comparison between protected text (mem-samples) and non-member samples compared to the No-Guard (NG).
    }
    \label{fig:demo}
\end{figure}

\subsubsection{Uniform Random Augmentation}\label{sec:uniform}
We start with a straw-man perturbation with uniformly random tokens.
Let $\x=\{x_1, x_2, \cdots, x_t\}$ denote the tokenized sequence with length of $t$, we randomly sample $m$ augmented random noise tokens as $\mathbf{\delta}=\{\delta_1, \delta_2, \cdots, \delta_m\} \in \mathcal{V}^{m}$ from a typical vocabulary set $\mathcal{V}$.
And $m=\lfloor b*t \rfloor$ is limited by the perturbation budget $b$. We split random tokens into $K$ pieces and insert them into $K$ slots within the original $\x$, where $K \leq t-1$.
We create two versions with different inserting positions:
\begin{itemize}
	\item \textbf{Uniform and Deterministic Perturbation (UDP)}: The token sequence is split into $K$ equal-length blocks, with $m$ random tokens inserted evenly.
	\item \textbf{Uniform and Non-deterministic Perturbation (UNP)}: $K$ slots are randomly chosen and filled with $m$ random tokens, which is a nondeterministic insertion.
\end{itemize}

\subsubsection{Verifying the Implicit Objective of Perturbation}\label{sec:verify}
Though the synthetic perturbation such as UDP and UNP does not optimize towards an explicit objective, we now demonstrate it essentially encourages the implicit objective of $\max\mathcal{L}_{\theta_\delta^*}(\x)$.

In \Cref{fig:demo} (Left) with UNP as an example, the gradient of protected tokens' embeddings change drastically compared to the case without perturbation (NG) across embedding indices (horizontal lines) and dimensions (vertical lines), which directly influences model update and results in \Cref{fig:demo} (Right).
With moderate perturbation ($b=1$), the influence on testing performance is trivial and the implicit target $\mathcal{L}_{\theta_\delta^*}(\x)$ increases, indicating that the target model is less likely to generate $\x$.

According to the analysis in \Cref{tab:proxy}, a larger loss on the target model $\mathcal{L}_{\theta_\delta^*}(\x)$ is expected to lower the general risk proxy of exploitation $\mathbf{Ex}(\x)$ as well as $\operatorname{Adv}(\x; \theta_\delta^*, \mathcal{A})$.
We will demonstrate both degradations in \Cref{sec:experiment}.

\subsection{Memorization Trigger Hypothesis}\label{sec:hypothesis}
To enhance defense efficacy, we first investigate how language models (LMs) memorize specific texts.
Unlike generalization, memorization occurs when a model captures sample-specific patterns rather than generalizable features.
From \Cref{eq:llm_loss}, we observe that rare or challenging tokens—those with higher initial loss values—leave a stronger imprint on the trained model compared to others.
This suggests that the model prioritizes memorizing these tokens over other easy tokens during training.
Thus, we hypothesize that:
\begin{hypothesisbox}
The model's memorization of input $\x$ primarily stems from its retention of \textit{hard-to-predict tokens} in $\x$---which we term \textbf{\textit{memorization triggers}}. 
\end{hypothesisbox}

\begin{figure}[t]
    \centering
    \begin{subfigure}[b]{0.47\linewidth}
        \includegraphics[width=\textwidth, trim=0 0 0 0, clip]{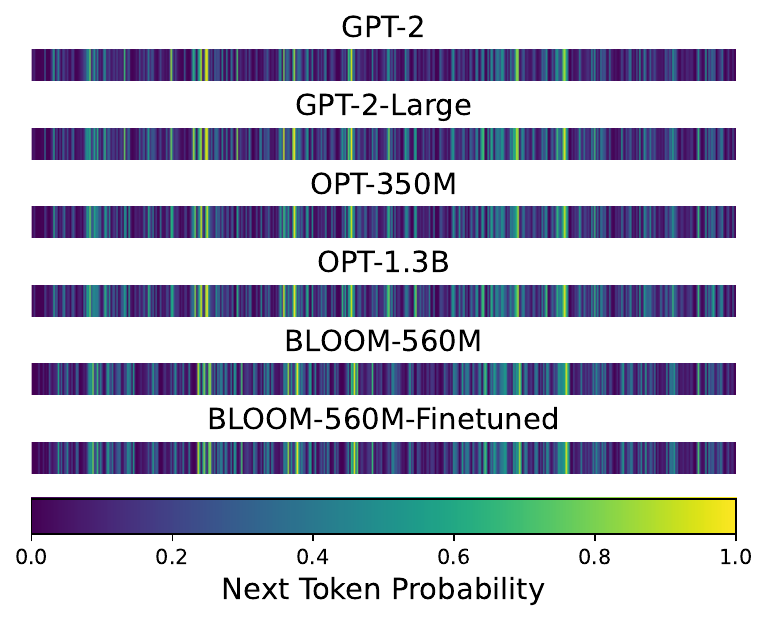}
        \caption{Transferable Identification}
        \label{fig:demo_mink_prob}
    \end{subfigure}
    \hfill
    \begin{subfigure}[b]{0.47\linewidth}
        \includegraphics[width=\textwidth, trim=0 0 0 20, clip]{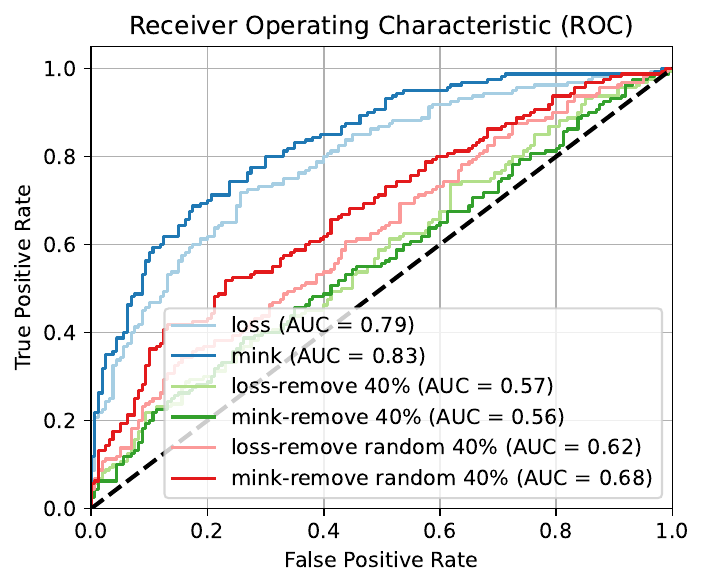}
        \caption{Attack-Defense Duality}
        \label{fig:demo_mink_remove}
    \end{subfigure}
    \caption{
    Memorization trigger hypothesis:
    tokens with higher loss on a proxy model are memorization triggers that enhance the sample-specific memorization.
    }
    \label{fig:demo_mink}
\end{figure}

\noindent \textbf{Transferable Identification.}
We define \textit{hard-to-predict tokens} as tokens assigned low probability by a general-domain pretrained model in their respective contexts.
In implementation, we leverage an open-source pre-trained model as a proxy $\theta_\text{proxy}$ and select tokens whose prediction probability $f_{\theta_\text{proxy}}(x_t|x_{<t})$ belong to the $K$ lowest-probability tokens in the sequence as a set $\mathbf{S}_{K}(\x)$.
\Cref{fig:demo_mink_prob} reveals that models across scales and families have a consistent identification on memorization triggers identification, indicating an architectural independence between proxy and target models, which is a key advantage for data owners acting as defenders.

\noindent \textbf{Attack-defense Duality.}
Then, we verify the existence of memorization triggers from both attack and defense perspectives with $K/|\x|=0.4$.
From the attack perspective, recent improvement on attack sheds similar light on the memorization trigger. 
As shown in \Cref{fig:demo_mink_remove}, MinK~\cite{shi2023detecting} outperforms loss-based MIA by only aggregating losses over low confidence tokens in the given sequence, implying that attackers rely on the improvement of the model's prediction capability on outlier tokens.
From the defense perspective, when memorization triggers are removed from the training data (as shown in \Cref{fig:demo_mink_remove}), the MIA AUC drops to near random-guess level ($\approx 0.56$). 
However, removing random tokens with identical ratio ($K/|\x|=0.4$) still maintains AUC 0.68, demonstrating the important role of memorization triggers in defense.

In summary, the memorization trigger hypothesis reveals the key mechanism for mitigating sample-specific memorization. In addition to being immediately applicable to proactive defense, this approach may also extend to later phases including pre-processing and training.

\subsection{Targeting on Memorization Triggers}
\label{sec:trigger}
Based on above hypothesis, a natural idea to enhance self-guard is to focus on perturbing identified memorization tokens.

\subsubsection{Targeted Perturbation}\label{sec:tp}
A simple extension is to insert random tokens right before the identified trigger tokens.
\begin{itemize}
	\item \textbf{Targeted Perturbation (TP):}
Instead of randomly sampling $K$ slots for inserting perturbation, we first identify Top-$K$ tokens in $\x$ with minimum prediction probabilities by a proxy model via $f_{\theta_\text{proxy}}(x_t|x_{<t})$.
Then, we insert uniform tokens as in UNP to fill slots before each of $K$ trigger tokens.
\end{itemize}

\subsubsection{Outlier Tokens as Pitfalls}\label{sec:pitfall}
Furthermore, instead of interfering model learning on the \textit{naturally} existed memorization triggers $\mathbf{S}_K(\x)$ as in TP, we propose to create \textit{artificial} memorization triggers to take the place of the original $\mathbf{S}_K(\x)$ as pitfalls.
By redirecting the model's optimization efforts toward these pitfall tokens, it mitigates model's memorization on the original $\mathbf{S}_K(\x)$.
Hence, we propose:
\begin{itemize}
	\item \textbf{Targeted Perturbation with Pitfalls (TP-P):} 
    After identifying memorization triggers $\mathbf{S}_K(\x)$, we feed preceding tokens before each slot at position $t$ to the proxy model $\theta_\text{proxy}$, and select the token $\arg \min_{v\in \mathcal{V}} f_{\theta_\text{proxy}}(v|x_{<t})$ as pitfall token to fill the slot iteratively until spending all budget $b$.
\end{itemize}

\subsubsection{Optimized Pitfalls}\label{sec:pitfall_opt}
From previous methods, we notice that the usage of the proxy model is insufficient, because the perturbation is sampled or generated.
Besides, instead of only considering the prefix of current token, we can also consider its context in the following variant.
\begin{itemize}
\item \textbf{Targeted Perturbation with Optimized Pitfalls (TP-OP)}: We first identify $K$ memorization triggers via the $\theta_\text{proxy}$. Then we optimize tokens to fill the position set $\mathcal{I}$ as follows.
\end{itemize}

Considering limited capabilities of defenders on training data and training algorithms as summarized in \Cref{tab:proxy}, we reformulate the bi-level optimization as single-level optimization by substituting the target model with the proxy model $\theta_\text{proxy}$ trained on general text.
By optimizing $\delta$ towards a maximized loss given  $\theta_\text{proxy}$, it creates a pitfall for the target model to fit during training. 
While on the contrary, a minimized loss can also help because the whole perturbed text is ignored by target model given the shortcut.
The two intuitions correspond to $\max$ and $\min$ cases in previous work~\cite{fowl2021adversarial}. \looseness=-1

We employ Greedy Coordinate Gradient (GCG)~\cite{zou2023universal} to optimize candidate tokens via the following objective, detailed in \Cref{alg:opt}:
\begin{align}
    \min_{\delta} \mathcal{L}_\text{P} &= \min_{\delta} [\beta_1 \mathbb{E}_{i\in\mathcal{I}}\mathcal{L}(\x_\delta[i];\x_{\delta}[<i], \theta_\text{proxy})  \nonumber \\
    &+ \beta_2 \mathbb{E}_{i\in\mathcal{I}} \mathcal{L}(\x_\delta[i+1]; \x_\delta[<i+1], \theta_\text{proxy}) ]\label{eq:proxy},
\end{align}
where coefficients $\beta_1, \beta_2\in\{1, -1\}$ represent error-minimization or error-maximization strategies.
By default, we use error-maximization with coefficient -1.

\begin{algorithm}[t]
\caption{TP-OP: Optimizing Pitfalls with GCG}\label{alg:opt}
\begin{algorithmic}[1]
\Require Iterations $\tau$, batch size $B$, number of token candidates $k$, position index set before $K$ memorization triggers $\mathcal{I}$, token vocabulary $\mathcal{V}$, batch size $B$
\Ensure Optimized pitfall tokens $\{x_{\delta, i}\}_{i \in \mathcal{I}}$
\Statex Randomly initialize inserted token embeddings $\{x_{\delta, i}\}_{i \in \mathcal{I}}$
\For{$j\in [\tau]$}
    \For{$i \in \mathcal{I}$} \textcolor{gray}{\textit{// Compute Top-$k$ promising candidates given the gradient of token embedding $e_{x_i}$ where $i\in |\mathcal{V}|$}}
        \State $\mathcal{X}_i := \text{Top-}k(-\nabla_{e_{x_i}} \mathcal{L}_\text{P}(\x_\delta))$ \label{line:topk}
    \EndFor
    
    \For{$b = 1, \ldots, B$} \textcolor{gray}{\textit{// Create a batch for searching}}
        \State $\tilde{x}_{1:m}^{(b)} := x_{1:m}$ \textcolor{gray}{\textit{// Initialize with current $n$ tokens}}
        \State $\tilde{x}_i^{(b)} := \text{Uniform}(\mathcal{X}_i)$, where $i = \text{Uniform}(\mathcal{I})$ 
    \EndFor
   
    \State $x_{1:m} := \tilde{x}_{1:m}^{(b^\star)}$, where $b^\star = \arg\min_b \mathcal{L}_\text{P}(\tilde{x}_{1:m}^{(b)})$
\EndFor
\end{algorithmic}
\end{algorithm}

\subsubsection{Out-of-Vocabulary (OOV) Tokens as Pitfalls}
\label{sec:pitfall_oov}
Previous methods perturb between tokens, leaving characters within original memorization triggers connected.
We propose an enhanced perturbation by breaking common tokens into out-of-vocabulary tokens that are harder to predict.
For example, inserting an invisible zero-width space in the word `language' as `lang\textbf{$\backslash$u200B}uage' completely transforms the original token sequence from [16129] to [17204, 9525, 84, 496] when using the GPT-2 tokenizer.
Given the set of all invisible characters $C$, we propose:\looseness=-1
\begin{itemize}
    \item \textbf{OOV-based Targeted Perturbation (TP-OOV)}, which first identifies memorization triggers as in \Cref{sec:trigger}, then randomly splits characters of each trigger token with a randomly sampled invisible character $c\in C$ within budget $b$. \looseness=-1
\end{itemize}
Additionally, we can combine this with \Cref{alg:opt} by replacing the candidate set $\mathcal{X}_i$ in Line 3 with set $C$, denoting this as TP-OOV++.
In practice, invisible characters should be filtered from both inputs and outputs in inference APIs for safety~\footnote{\url{https://www.promptfoo.dev/docs/red-team/plugins/ascii-smuggling/}}, which actually favors our defense since the artificial memorization triggers of OOV tokens will never be triggered during inference. \looseness=-1

\section{Experiments}\label{sec:experiment}

\begin{table*}[t]
\caption{Membership inference evaluation with maximum metrics reported for across Loss-Ref, Loss, Min-K, Zlib due to space limit, leaving full results in Appendix.
TPR is calculated at 1\%FPR.
BD indicates that the pre-trained model is backdoored to maximize the privacy risk of training data, i.e., BD represents the worst privacy leakage case.
}
\label{tab:main}
\centering

\begin{tabularx}{\linewidth}{c|c|*{6}{YY}} %
\toprule
\multirow{2}{*}{MIA Level} & 
  \multirow{2}{*}{Method} & 
  \multicolumn{2}{c|}{Patient} & 
  \multicolumn{2}{c|}{Enron} & 
  \multicolumn{2}{c|}{Patient} & 
  \multicolumn{2}{c|}{CC-News} &
  \multicolumn{2}{c|}{Patient} & \multicolumn{2}{c}{IAPR-TC-12}
  \\
 & & 
  \multicolumn{2}{c|}{GPT-2} & 
  \multicolumn{2}{c|}{OPT-350M} & 
  \multicolumn{2}{c|}{GPT-2 w/ BD} & 
  \multicolumn{2}{c|}{OPT-125M w/ BD} &
  \multicolumn{2}{c|}{Llama2-7B} &
  \multicolumn{2}{c}{BLIP2-ViT-3.8B}
  \\
  \cmidrule(lr){3-4} \cmidrule(lr){5-6} \cmidrule(lr){7-8} \cmidrule(lr){9-10}
  \cmidrule(lr){11-12} \cmidrule(lr){13-14} 
                        &                & AUC   & TPR & AUC   & TPR & AUC   & TPR & AUC   & TPR & AUC & TPR & AUC & TPR \\
\midrule
\multirow{7}{*}{Sample} & NP             & \cellcolor{gray!20}0.888 & \cellcolor{gray!20}0.364      & \cellcolor{gray!20}0.997 & \cellcolor{gray!20}0.983      & \cellcolor{gray!20}0.953 & \cellcolor{gray!20}0.545      & \cellcolor{gray!20}0.998 & \cellcolor{gray!20}0.982     & \cellcolor{gray!20}0.986 & \cellcolor{gray!20}0.726 & \cellcolor{gray!20}1.000 & \cellcolor{gray!20}0.980   \\
                        & UDP (b=0.4)    & 0.771 & 0.242      & 0.994 & 0.950      & 0.831 & 0.182      & 0.997 & 0.970     & 0.861	& 0.260 & 0.984 & 0.510 \\
                        & UNP (b=0.4)    & 0.695 & 0.182      & 0.986 & 0.735      & 0.766 & 0.152      & 0.983 & 0.467     & 0.852	& 0.164 & 0.984	& 0.560 \\
                        & TP (b=0.4)     & 0.686 & 0.182      & 0.979 & 0.621      & 0.765 & 0.182      & 0.978 & 0.580     & 0.856	& 0.219 & \textbf{0.509} & 0.010 \\
                        & TP-P (b=0.4)   & 0.682 & 0.212      & 0.989 & 0.837      & 0.772 & 0.182      & 0.991 & 0.746     & 0.793	& 0.123 & 0.550	& \textbf{0.000} \\
                        & TP-OOV (b=0.4) & 0.594 & 0.091      & 0.892 & 0.254      & 0.587 & 0.091      & 0.890 & 0.083     & 0.753	& 0.082 & 0.551	& \textbf{0.000} \\
                        & TP-OOV (b=1)   & \textbf{0.590} & \textbf{0.060}      & \textbf{0.684} & \textbf{0.119}      & \textbf{0.550} & \textbf{0.076}      & \textbf{0.621} & \textbf{0.053}     & \textbf{0.630}	& \textbf{0.055} & 0.519	& 0.010 \\
\midrule
\multirow{7}{*}{User}   & NP             & \cellcolor{gray!20}0.676 & \cellcolor{gray!20}0.047      & \cellcolor{gray!20}0.987 & \cellcolor{gray!20}0.585      & \cellcolor{gray!20}0.741 & \cellcolor{gray!20}0.047      & \cellcolor{gray!20}0.966 & \cellcolor{gray!20}0.035     & \cellcolor{gray!20}0.936	& \cellcolor{gray!20}0.452	& \cellcolor{gray!20}0.974	& \cellcolor{gray!20}0.377 \\
                        & UDP (b=0.4)    & 0.617 & 0.039      & 0.968 & 0.439      & 0.649 & 0.047      & 0.948 & 0.035     & 0.749	& 0.096	& 0.901	& 0.057 \\
                        & UNP (b=0.4)    & 0.598 & 0.039      & 0.933 & 0.269      & 0.622 & 0.039      & 0.912 & \textbf{0.032}     & 0.740	& 0.082	& 0.907	& 0.140 \\
                        & TP (b=0.4)     & 0.584 & 0.039      & 0.921 & 0.219      & 0.618 & 0.039      & 0.918 & 0.035     & 0.746	& 0.082	& \textbf{0.511}	& \textbf{0.003} \\
                        & TP-P (b=0.4)   & 0.588 & 0.039      & 0.951 & 0.282      & 0.619 & 0.039      & 0.923 & 0.035     & 0.667	& 0.068	& 0.541	& 0.007 \\
                        & TP-OOV (b=0.4) & 0.539 & 0.039      & 0.777 & 0.123      & \textbf{0.542} & 0.039      & 0.783 & 0.035     & 0.682	& 0.082	& 0.535	& \textbf{0.003} \\
                        & TP-OOV (b=1)   & \textbf{0.567} & \textbf{0.031}      & \textbf{0.640} & \textbf{0.090}      & 0.567 & \textbf{0.031}      & \textbf{0.605} & 0.035     & \textbf{0.545}	& \textbf{0.041}	& 0.523	& \textbf{0.003} \\
\bottomrule
\end{tabularx}
\end{table*}

\subsection{Experimental Setup}
\noindent\textbf{Tasks and Datasets.}
We conduct comprehensive evaluation on general language modeling and vision-to-language modeling (VLM) tasks.
We include representative data sources requiring protection but potentially disclosed in web content:
1) Enron~\cite{klimt2004enron} with personal information (emails, names, medical records), 
2) Patient~\cite{zeng2020meddialog} with domain knowledge (healthcare), 
3) CC-News~\cite{mackenzie2020cc} with copyrighted work (news articles), and 
4) IAPR-TC-12~\cite{iapr} with natural images and descriptions for VLM tasks.
Dataset details are in \Cref{sec:data-APP}.

For risk evaluation, we split the dataset at a ratio 1:4:4:1, with $D$ for training the target model, $D_\text{aux}$ for reference model or privacy backdoor, $D_\text{non}$ as non-members,  and $D_\text{test}$ for testing.
We split $D$ by marking a fraction $r \in (0,1]$ as protected ($D_\text{pro}$) and the rest as unprotected ($D_\text{un}$).

\noindent\textbf{Models and Training Configuration.}
We evaluate on open-source models due to our requirement for per-sample losses and token-level probabilities—information unavailable through commercial APIs.
We use GPT-2 as the proxy model for all tasks, showing architecture independence.
We evaluate on models of different scales: GPT-2-124M~\cite{radford2019language}, OPT-125M/350M~\cite{zhang2022opt}, Llama2-7B~\cite{touvron2023llama} for LM, and BLIP-2-ViT (3.8B) for VLM.
The VLM model processes image inputs and autoregressively generates text conditioned on preceding inputs.

\noindent\textbf{Evaluation Configuration.}
For \textit{individual-level} risk evaluation, we use the exploitation metric defined in \Cref{def:exploit}, and we approximate $\theta_{\backslash \x}$ with $\theta_{\backslash D_\text{pro}}$ as in \Cref{eq:exp_approx}.

For \textit{dataset-level} risk evaluation, we cover practical attacks including data extraction and membership inference attacks (MIA).
For MIA, we evaluate using state-of-the-art threshold-based black-box MIAs with signals: 
1) Loss~\cite{yeomPrivacyRiskMachine2018} (model's loss on target samples), 
2) Loss-Ref~\cite{carlini2021extracting} (loss calibrated against reference model), 
3) MinK~\cite{shi2023detecting} ($K$\% tokens with lowest likelihood scores), and 
4) Zlib~\cite{carlini2021extracting} (loss normalized by compression size).
We use `user-level' (each sample belongs to one user) and `sample-level' (documents chunked to full window size) evaluation, with the sample index as the user ID, except for Enron, which has explicit user IDs.
For data extraction, we follow the recent work~\cite{hayes2024measuring} to evaluate success of extracting $T$-length sub-sequences in $D_\text{pro}$ over $N$ trials.

For the \textit{worst-case} risk evaluation, we follow recent works and assume a malicious and powerful model trainer who can manipulate a significant portion of $D_{\backslash \x}$ to insert privacy backdoor~\cite{liu2024precurious, wen2024privacy} with details in \Cref{sec:backdoor-game-APP}.

\noindent\textbf{Evaluation Metrics.}
Following previous work~\cite{carlini2022membership}, we measure MIA risk with AUC$\downarrow$, true positive rate at low false positive rate (TPR@1\% FPR$\downarrow$) and ROC curves. We report bootstrapped metrics~\cite{efron1992bootstrap} for stability.
At the individual-level, we report approximated exploitation $\hat{\mathbf{Ex}}(\x)\downarrow$ for each sample.
Lower values indicate lower privacy risk.
For utility cost, we report perplexity (PPL$\downarrow$) on held-out test data, reflecting the implicit cost to model trainers since data owners are not obligated to maintain high performance.

\noindent\textbf{Baselines.} Supposing the model trainer is neutral and not motivated to perform training-phase defense, we compare \ours with the baseline of 
No Protection (NP) where users release their text contents without any protection.
To fully investigate each proposed strategy, we evaluate all our variants in \Cref{tab:ours}, with moderate perturbation budget $b$.

\subsection{Effectiveness Evaluation}
\textbf{Effectiveness against MIAs.} 
As shown in \Cref{tab:main}, we compare MIA risks across different tasks and datasets.
The gap between NP and our variants demonstrates overall defense effectiveness under identical training and attack pipelines for $D_\text{pro}$.
Sample-level evaluation exhibits higher MIA metrics than user-level evaluation due to longer chunk lengths~\cite{puerto2024scaling}.
We intentionally train models with slight overfitting—large models with billions of parameters (Llama2-7B and BLIP2-ViT-3.8B) achieve an AUC of $\approx 1$ for NP, representing the worst-case scenario for defenders.

Comparing UDP and UNP results confirms that non-deterministic perturbation outperforms the deterministic variant, as models can learn to ignore deterministic patterns.

We observe consistent TP improvements over UNP across different architectures, including Enron results where proxy and target models differ, confirming the model transferability demonstrated in \Cref{fig:demo_mink_prob} of \Cref{sec:hypothesis}.
Notably, transferability performs better for memorization identification than pitfall creation—TP-P shows slightly higher risk than TP for Enron.
When proxy and target models share identical architectures (Patient dataset), replacing random tokens in TP with outlier pitfalls in TP-P yields only marginal improvements.

With perturbation budget $b=0.4$, TP-OOV achieves superior defense effectiveness among all variants, primarily due to memorization trigger identification and out-of-vocabulary sub-tokens.
On GPT2 and BLIP2, TP-OOV with sufficient budget $b=1$ approaches random performance, indicating that GPT-2 tokenizer-generated OOV pitfalls generalize across different target models.
This effectiveness stems from shared lexical knowledge across language models, and tokens that are out-of-vocabulary in one LM typically remain so in others.

\begin{figure}
    \centering
    \includegraphics[width=0.45\linewidth]{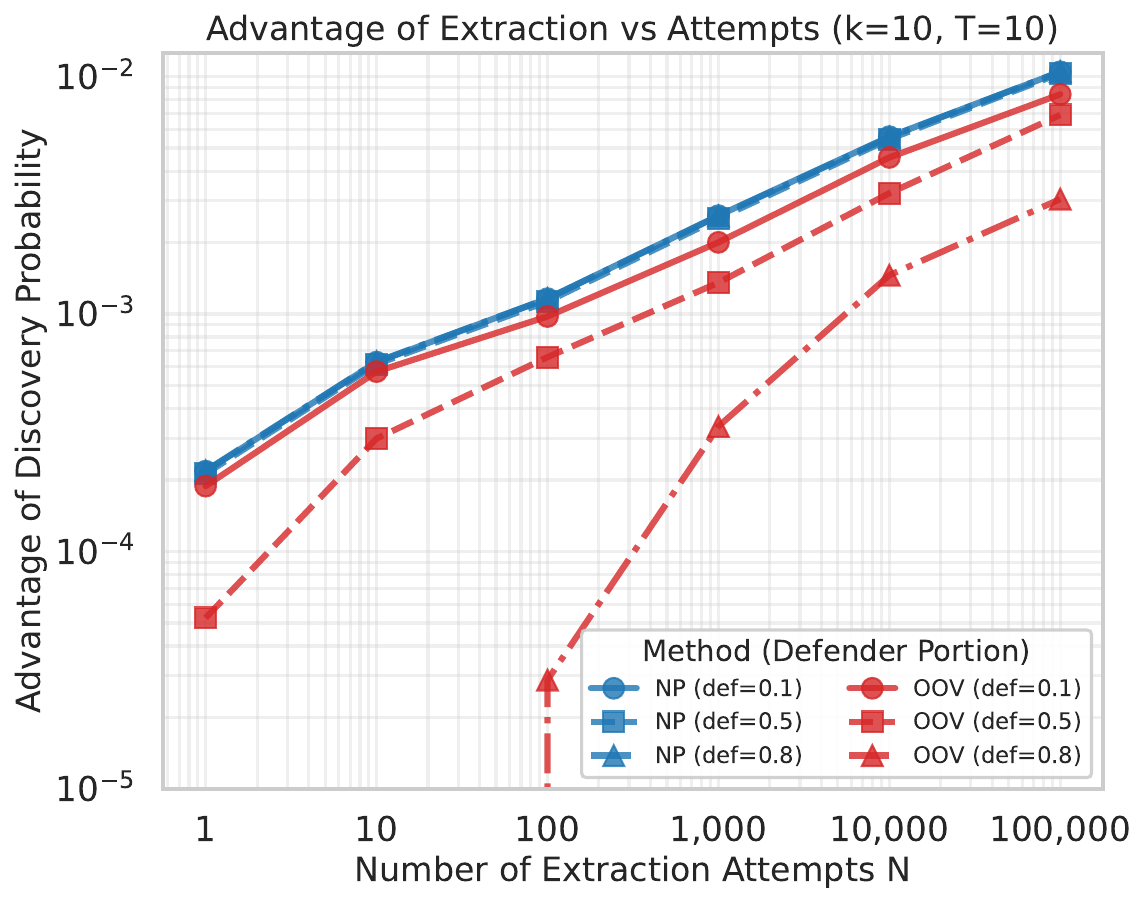}
    \includegraphics[width=0.45\linewidth]{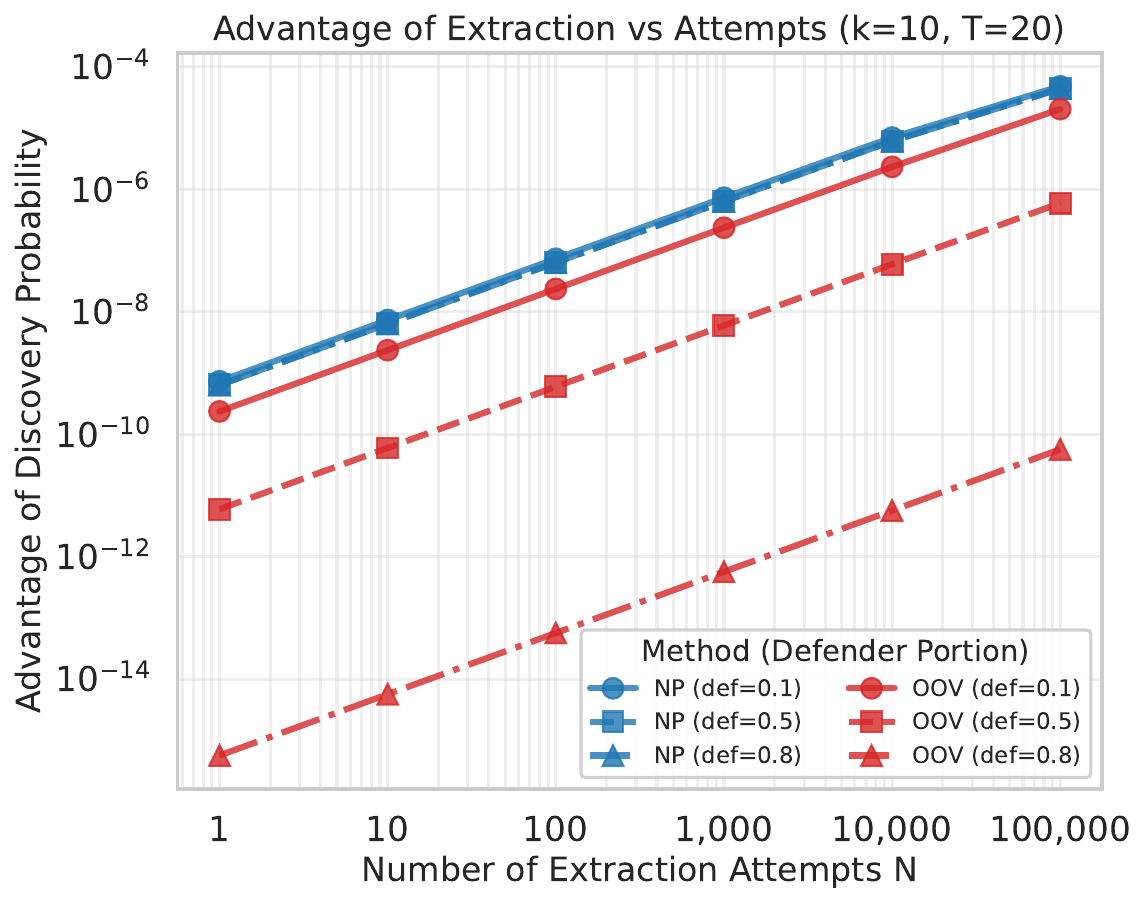}
    \includegraphics[width=0.45\linewidth]{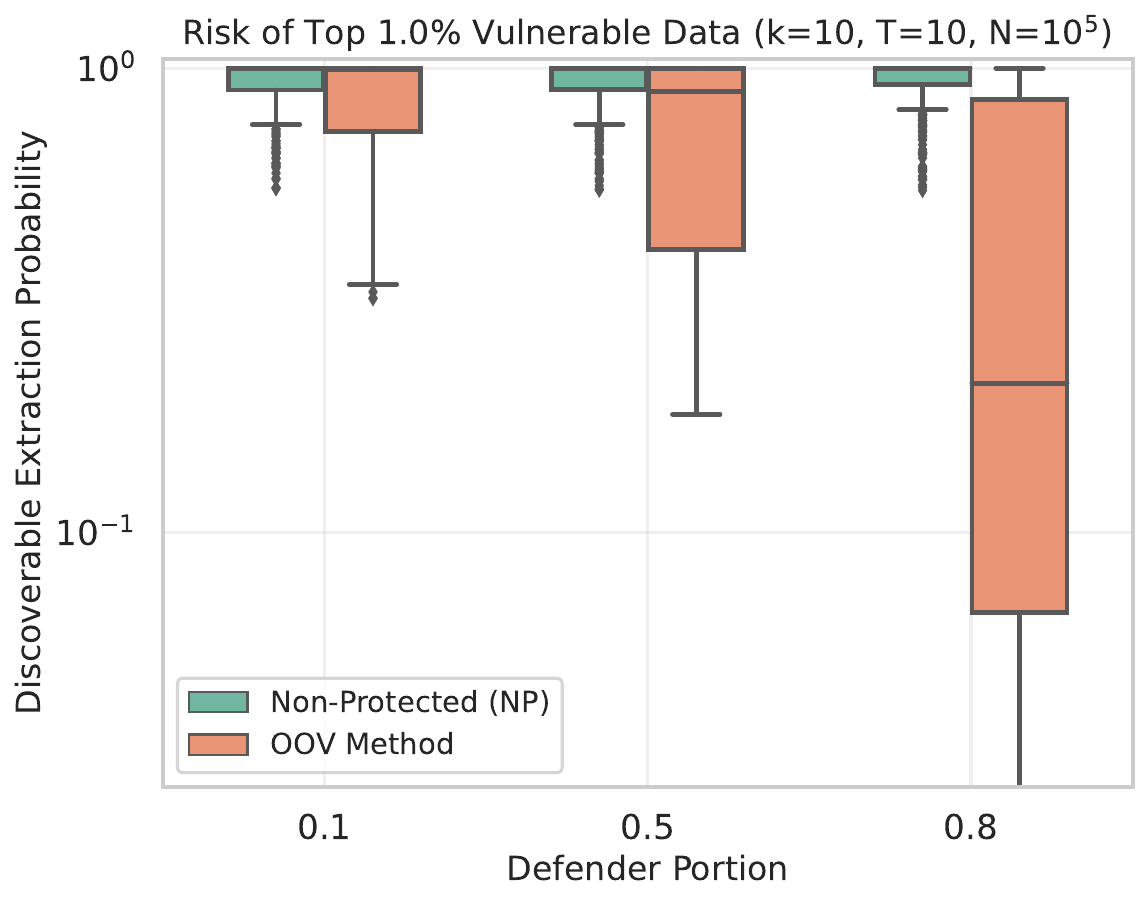}
    \includegraphics[width=0.45\linewidth]{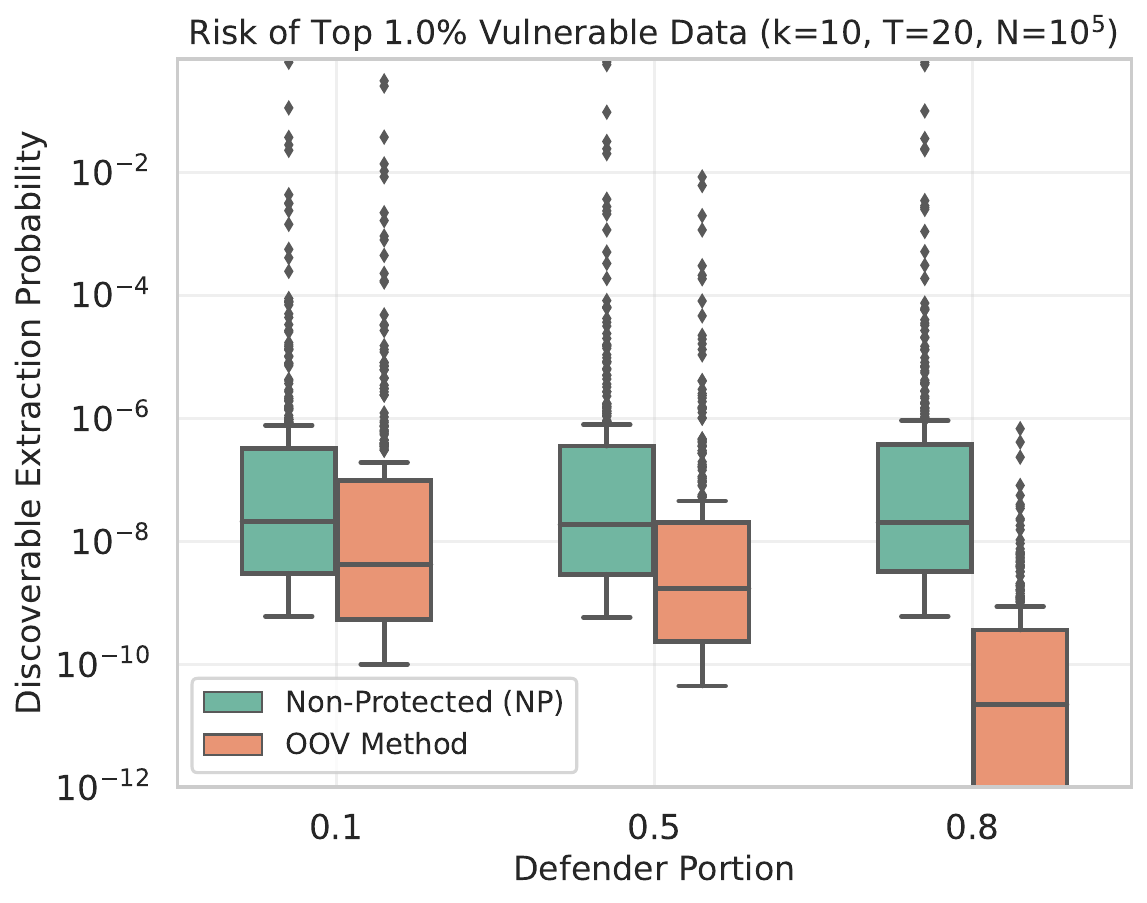}
    \caption{Discoverable Data Extraction Evaluation.}
    \label{fig:ext_top}
\end{figure}

\noindent\textbf{Effectiveness against Data Extraction.}
In \Cref{fig:ext_top}, we evaluate discoverable data extraction~\cite{hayes2024measuring} risk using TP-OOV against unprotected baseline NP.
We model realistic adversaries performing $N=10^5$ extraction attempts on subsequences of length $T\in\{10,20\}$ using Top-$k$ decoding ($k=10$), which sets a loose upper bound on the adversary's access capability.
The extraction probability quantifies the likelihood that the target model generates the protected subsequence within $N$ attempts.
The extraction advantage measures how much more information the target model provides compared to a GPT-2 proxy model about protected subsequences.

Results in the top row demonstrate that TP-OOV consistently reduces extraction advantage across all $N$, $T$, and protection ratios $r$.
When advantage approaches zero as the case for $r=0.8, N<100$, the target model provides no more information than the proxy model.
Defense effectiveness is particularly strong for shorter sequences ($T=10$) and scales with protection coverage—larger $r$ values yield greater risk reduction due to reduced overlap with unprotected content.
In the bottom figures, analyzing the Top-1\% most vulnerable subsequences confirms a statistically risk degradation.

\noindent\textbf{Effectiveness under Privacy Backdoor.}
Furthermore, as an empirical defense, it is necessary to evaluate it under the current most powerful privacy attacks.
Therefore, we apply the recent privacy backdoor~\cite{liu2024precurious, wen2024privacy} to amplify the privacy risk of fine-tuning training data by assuming that the pre-trained model is released and crafted by the privacy attacker.
Specifically, we warm up the pre-trained backbone on $D_\text{aux}$ for 2 more epochs before fine-tuning, which pushes the model to enter the memorization-only stage earlier than using a benign pre-trained model and to memorize more unique details of its training samples.
Additionally, we use the warmed-up pre-trained model as the reference model in Loss-Ref MIA.

As shown in \Cref{tab:main}, comparing the privacy risk of Patient dataset with the same attack and training setting, the
privacy risk is indeed amplified.
Similarly, for CC-News dataset on OPT-125M model, the AUC and TPR approach 1 for Loss-Ref and MinK, indicating that the attack is near perfect.
Nonetheless, by applying TP-OOV with $b=1$, \ours reduces the maximum TPR at 1\% FPR across MIA signals from 0.982 to 0.053.
In general, the MIA evaluation demonstrates that \ours successfully offers protection to the data owner by reducing the overall privacy risk of $D_\text{pro}$ even when the MIA is near perfect.

\begin{figure}[t]
    \centering
    \begin{subfigure}[b]{0.48\linewidth}
        \includegraphics[trim=8 8 5 20, clip, width=\linewidth]{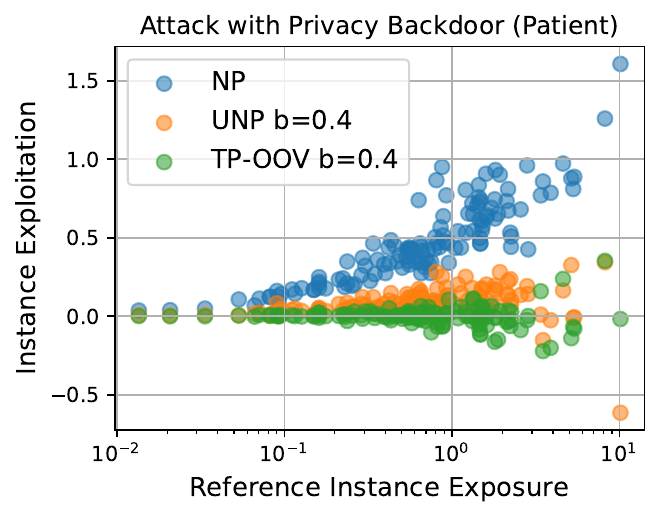}
        \caption{LM Patient w/ DB}
        \label{fig:instance}
    \end{subfigure}
    \hfill
    \begin{subfigure}[b]{0.48\linewidth}
        \includegraphics[width=\linewidth]{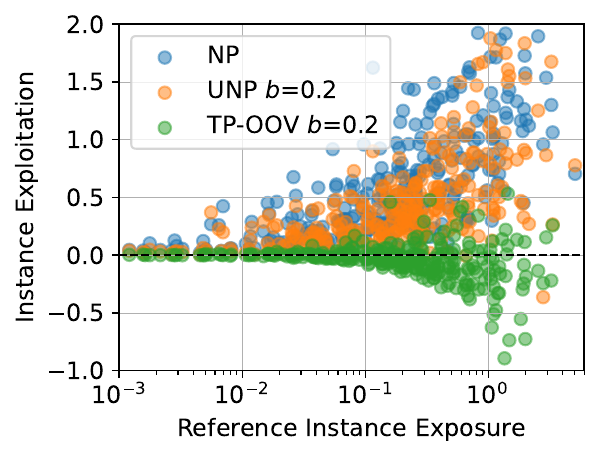}
        \caption{VLM IAPR-TC-12}
        \label{fig:ins_vlm}
    \end{subfigure}
    \caption{Instance vulnerability of representative variants.
    }
    \label{fig:combined_analysis}
\end{figure}

\begin{figure}
    \centering
    \includegraphics[trim=0 0 0 25, clip, width=0.48\linewidth]{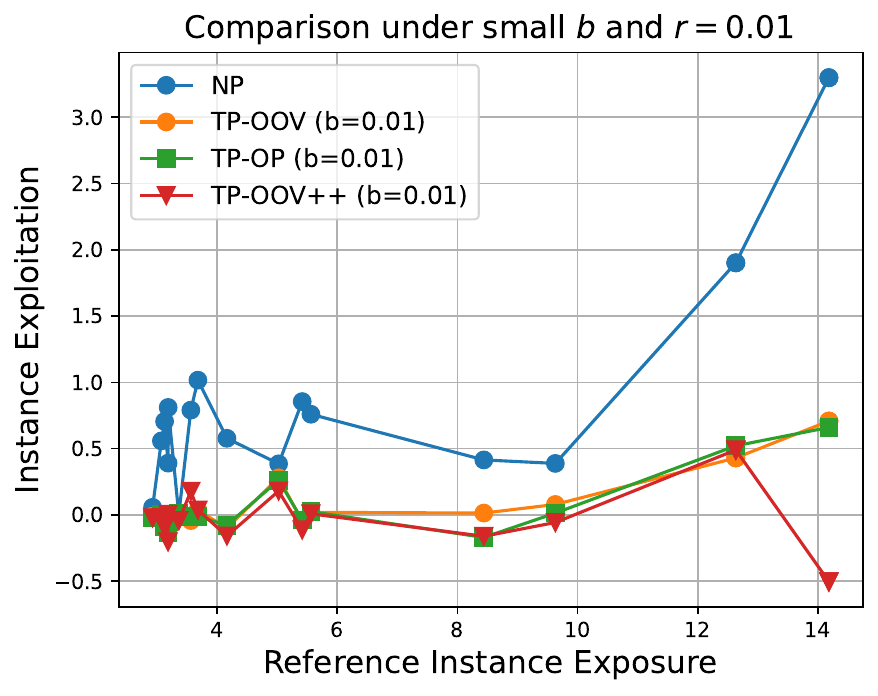}
    \includegraphics[trim=0 0 0 25, clip, width=0.47\linewidth]{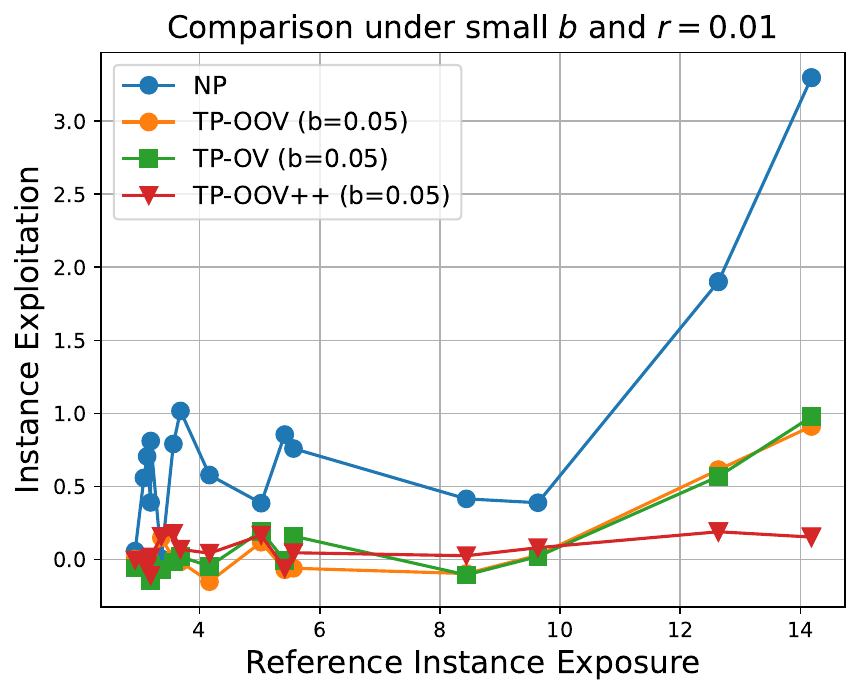}
    \caption{Effectiveness of optimization-based method on most vulnerable instances given a small portion of defender $r=0.01$ and small portion of perturbation budget.}
    \label{fig:eff_opt}
\end{figure}

\noindent\textbf{Instance-Level Risk Evaluation}.
Beyond the averaged risk over $D_\text{pro}$, we evaluate individual-level risk for vulnerable instances using the proposed instance exploitation defined in \Cref{def:exploit}.
As shown in \Cref{fig:instance}, we compare the instance exploitation for each sample (corresponding to each point) in $D_\text{pro}$ as a function of the sample $\x$'s instance exposure $E_{\theta\backslash \x}(\x)$ obtained from a model not trained on it.

We first identify a general pattern in the unprotected baseline (NP) across different datasets and models: The sample that is originally more exposed than other samples (with a higher $E_{\theta\backslash \x}(\x)$) typically has a higher instance exploitation after the model is trained on the sample.
In other words, naturally vulnerable instances are prone to being more exposed in future training, which aligns with our memorization trigger hypothesis.
And the reason is that the training objective is designed to focus on samples with higher loss.
CC-News and Enron exhibit analogous trends (Appendix \Cref{fig:instance-APP}).

As two representative methods without a proxy model (UNP) and with proxy model (TP-OOV), we observe that the instance exposure even for the naturally exposed samples is significantly reduced.
Furthermore, by leveraging the proxy model to identify memorization triggers and creating pitfalls in TP-OOV, all protected instances have instance exploitation below 0.5, meaning that \ours prevents the protected instance from being exploited beyond the general knowledge in training distribution.\looseness=-1

\begin{figure}
\centering
	\includegraphics[trim=0 0 0 0, clip, width=0.5\textwidth]{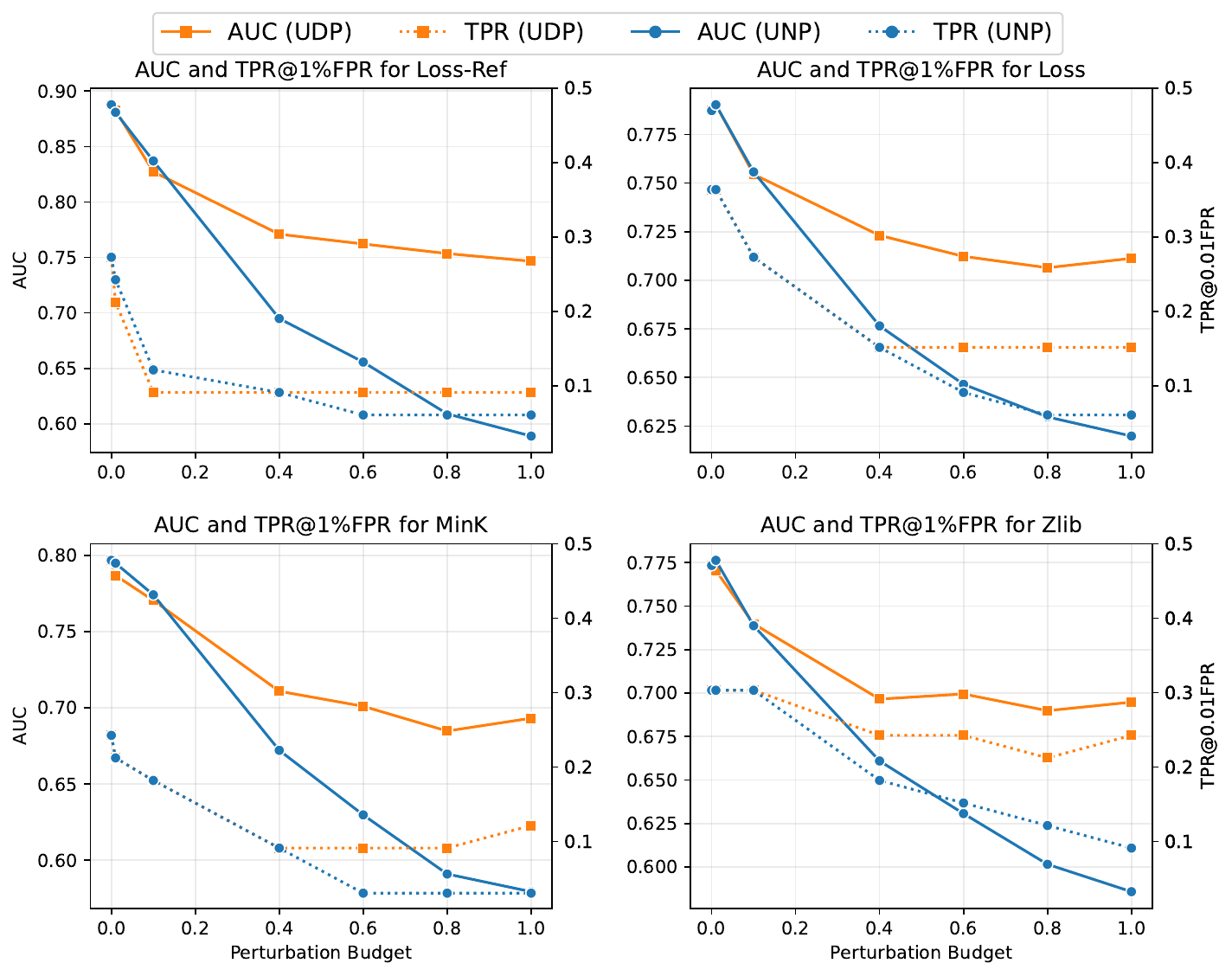}
	\caption{Disturbing strategy evaluation on deterministic perturbation (UDP) and non-deterministic perturbation (UNP).
	}\label{fig:inf_disturb}
\end{figure}

\noindent\textbf{Extending to Vision-Language Modeling.}
In \Cref{tab:main}, we can observe that perturbing memorization triggers (TP) significantly reduces the privacy risk from perfect attack to near random guess with AUC around 0.509.
In \Cref{fig:ins_vlm}, we observe a similar trend as in LM that more vulnerable samples which have a higher reference instance exposure tend to have higher instance exploitation for NP.
Even for those vulnerable samples, the instance exploitation of each text sample approaches zero with a small budget $b=0.2$ for TP-OOV, indicating a perfect individual defense for VLM.
Due to the space limit, we present the details and additional results in \Cref{app:vlm}.

\noindent\textbf{Effectiveness of Optimization.}
We now evaluate the effectiveness of optimization-based method with variant TP-OP and the extended version TP-OOV++ for TP-OOV by integrating it with the optimization based method as discussed in \Cref{sec:pitfall_oov}.
We note that optimizing for one sentence is affordable, but it is time-consuming to perform the optimization over every instance in $D_\text{pro}$.
Since we have shown that instances having a naturally high exposure are more prone to having high exploitation, we focus on the most vulnerable data points when we evaluate the optimization-based extension.
Thus, we first select a portion $r=0.1$ as $D_\text{pro}$ and use the rest of the unprotected samples to train a reference model $\theta_{\backslash\x}$.
Then we calculate $Ex_{\theta{\backslash}}(\x)$ for every $\x \in D_\text{pro}$ and only keep the vulnerable subset with the Top-20 highest exposure instances in $D_\text{pro}$ for perturbation.
After training on $D_\text{pro}\cap D_\text{un}$, we can obtain the instance exploitation as shown in \Cref{fig:eff_opt}.

We can observe that optimizing over inserted tokens (TP-OP) has similar effectiveness to breaking up the top memorization triggers (TP-OOV). In addition, extending TP-OOV by searching for an OOV that has a maximized loss via  $\beta_1=-1$ further reduces the instance exploitation for the highest-exposed instance.
We leave similar results under different $b$ in \Cref{APP:sec:exploitation}.

\subsection{Hyper-Parameter and Ablation Analysis}\label{sec:hyper}
We conduct detailed hyperparameters and ablation analysis to understand the rationale of different variants of \ours.

\noindent\textbf{Influence of Disturbing Pattern.}
We show that the variant of non-deterministic perturbation (UNP) is superior to deterministic perturbation (UDP) as shown in \Cref{fig:inf_disturb}, especially when $b$ is larger, because the model can learn how to ignore the perturbation when there is a deterministic pattern.

\begin{figure}
\centering
	\includegraphics[trim=0 0 0 0, clip, width=0.5\textwidth]{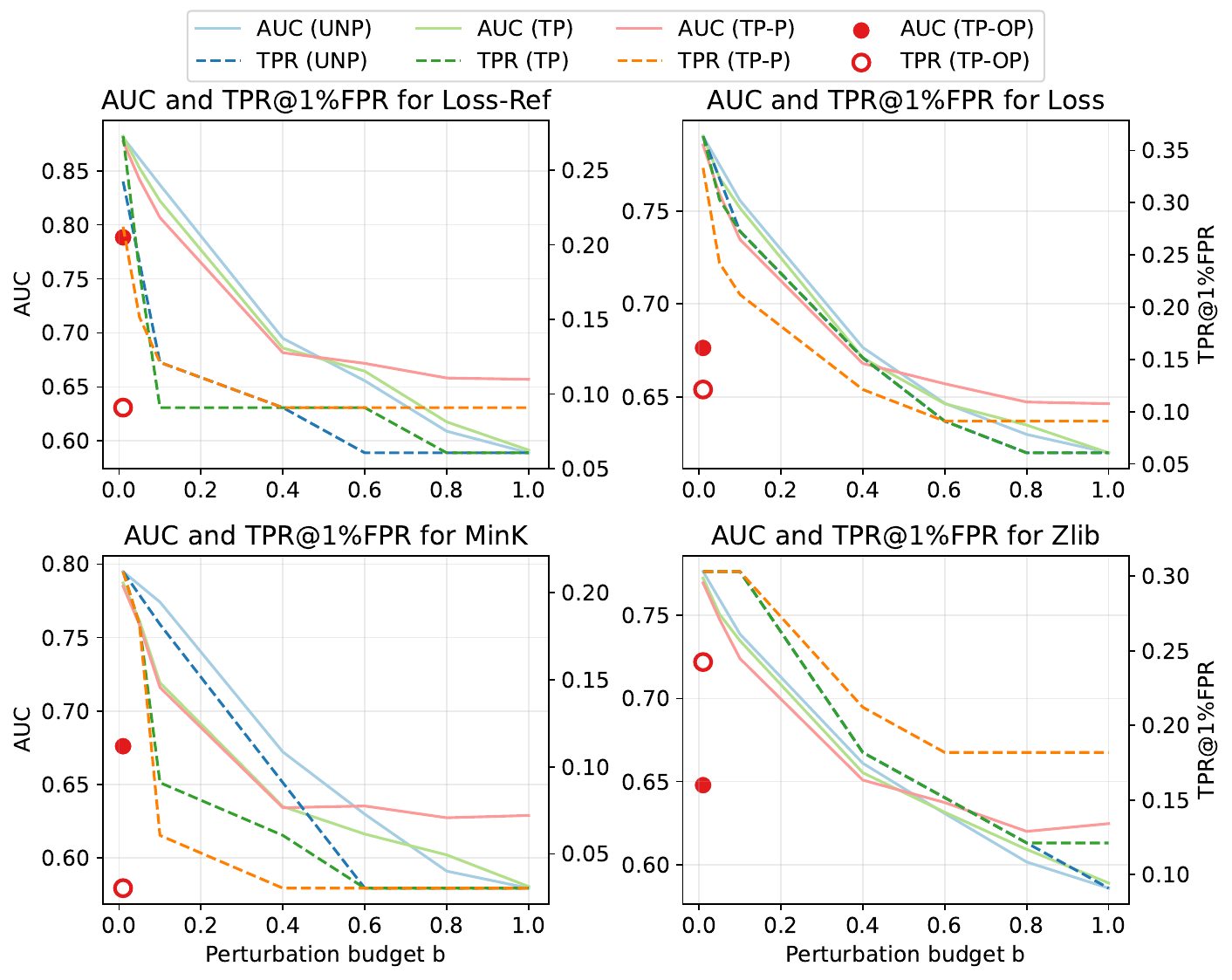}
	\caption{Evaluation of token-level filling strategies on uniformly random tokens (TP), pitfall tokens (TP-P) and optimized pitfall tokens (TP-OP).
	}\label{fig:inf_filling}
\end{figure}

\begin{table}[thb]
\centering
\caption{Hyper-parameter analysis for TP-OP. Optimal metrics are bold and underlines indicate improvement over random perturbation in UNP.}
\label{tab:beta}
\resizebox{0.5\textwidth}{!}{
\begin{tabular}{c|cc|cc}
\toprule
\multirow{2}{*}{Method} & \multicolumn{2}{c}{Sample-level}            & \multicolumn{2}{c}{User-level}              \\
                         & Max-AUC     & Max-TPR@1\%              & Max-AUC     & Max-TPR@1\%              \\
                         \midrule
NP                       & 0.888       & 0.364                & 0.676       & 0.047                \\
\midrule
UNP                      & 0.881       & 0.364                & 0.674       & 0.039                \\
\midrule
$\beta_1=1, \beta_2=0$   & \underline{0.837}       & \underline{0.303}                & \underline{0.646} & \underline{\textbf{0.039}} \\
$\beta_1=0, \beta_2=1$  & \underline{\textbf{0.782}} & \underline{\textbf{0.242}} & \underline{\textbf{0.619}} & 0.070                \\
$\beta_1=1, \beta_2=1$   & \underline{0.788} & \underline{\textbf{0.242}} & \underline{0.624} & 0.070                \\
$\beta_1=-1, \beta_2=1$ & \underline{0.784}          & \underline{0.273}          & \underline{0.623}          & \underline{\textbf{0.039}} \\
$\beta_1=-1, \beta_2=-1$ & \underline{0.788} & \underline{\textbf{0.242}} & \underline{0.632} & 0.047            \\
\bottomrule
\end{tabular}
}
\end{table}

\noindent\textbf{Influence of Filling Strategies.}
We compare different filling strategies in \Cref{fig:inf_filling}.
We observe that filling with pitfall tokens (TP-P) enhances the defense against most MIAs especially when the perturbation budget $b<0.6$.
And filling with optimization-based pitfall tokens is more effective than other variants on decreasing the privacy risk across all signals, under the perturbation constraint of $b=0.01$.

\noindent\textbf{Influence of $\beta_1$ and $\beta_2$.}
We then analyze the choice of $\beta_1$ and $\beta_2$ in \Cref{eq:proxy} where a positive coefficient indicates error-minimization and a negative coefficient denotes error-maximization.
As shown in \Cref{tab:beta}, in general we observe that either a negative or positive coefficient helps to reduce privacy risk.
Setting $\beta_1=1$ is less effective, as we are optimizing the perturbation to make it as fluent as possible given previous context, which makes the target model ignore the inserted token.
On the contrary, $\beta_2=1$ yields lower risk, as we encourage the connection between the inserted tokens and the identified memorization triggers, which creates a dependency from memorization-triggered tokens on the inserted token and fools the model to focus on learning  perturbation.

\begin{figure*}
    \centering
    \includegraphics[width=0.96\linewidth]{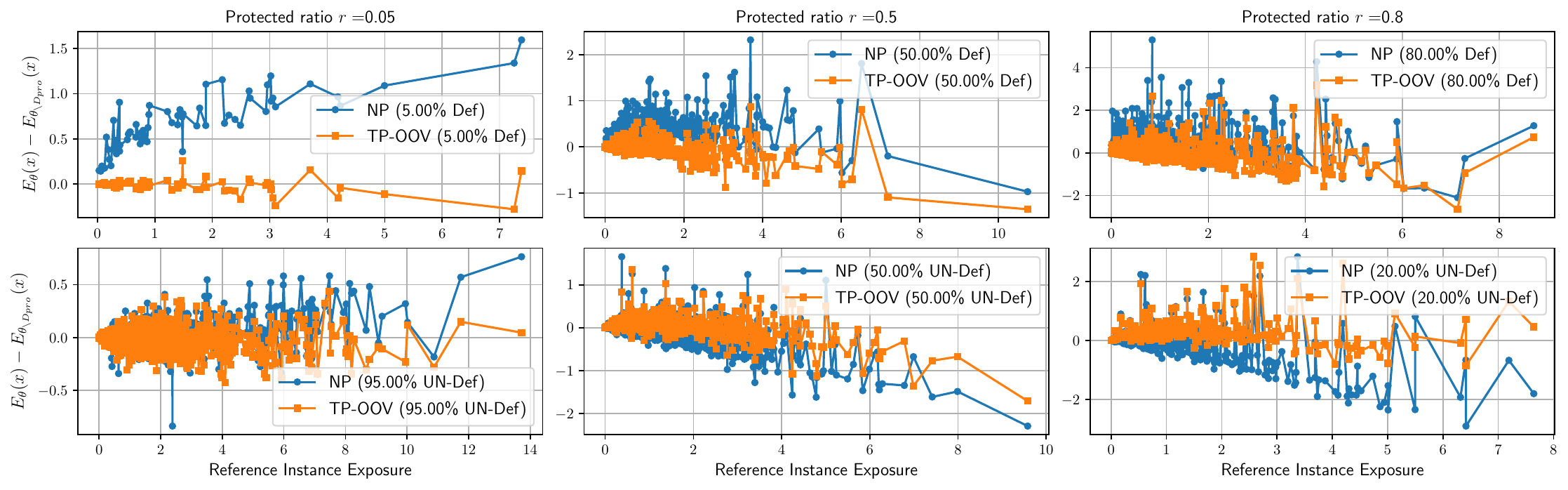}
    \caption{The influence defender ratio $r$ on the individual risk of other unprotected instances.
    For figures in each column, we train the exposure reference model with $D_{\backslash D_\text{pro}}$ separately for approaching training on $D_{\backslash \x}$ as reference, while larger $r$ results in a looser approximation and the gap becomes smaller from left to right.
    The y-axis indicates the exposure change after introducing $D_\text{pro}$.
    The gap between NP and TP-OOV demonstrates the exposure change caused by applying \ours.
    }
    \label{fig:inf_r_un}
\end{figure*}

\noindent\textbf{Influence on Model Utility.}
We also compare the model utility with the initial pre-trained model and the model trained without any protection (NP) in \Cref{tab:utility}.
Even when the defender portion ($r=0.8$) and perturbation budget ($b=1$)  are large, the validation performance for a model trained on TP-OOV drops compared to NP. It is still  significantly better than the initial model, indicating that the model is able to learn from the unprotected data.  
The `Mem-Loss' is the average loss of all protected instances, which remains high at 4.023 while the `Mem-Loss' for NP baseline decreases a lot to 3.586, indicating that the model does not learn much information from $D_\text{pro}$.
We also notice that there is a significant decrease in the training loss to 2.351, which means the model learns from the combination of unprotected and protected samples rapidly.
This is because inserted pitfalls are designed to be outliers in $D$, and the model is prone to focus on this pattern.

In \Cref{tab:utility_out}, we present the utility degradation for OUT, which removes all protected instances from $D$, and for TP-OOV, given different defender ratio $r$ with budget $b=1$ for Patient dataset.
We observe that self-guarding with \ours achieves a slightly higher utility drop than filtering out protected samples directly.
But when all samples are self-guarded and the perturbation against reference-model-based MIA is strong, it is possible that the model's validation performance is worse than the initial pre-trained model.
While in a more realistic setting where only the minority of web content is self-guarded ($r<0.1$), the utility loss compared to OUT is small.

\begin{table}[t]
\centering
\caption{Training and validation performance.}
\label{tab:utility}
\resizebox{0.4\textwidth}{!}{
\begin{tabular}{c|cc|cc}
\toprule
Method         & Val-PPL & Val-Loss & Train-Loss & Mem-Loss \\
\midrule
Initial model  & 65.412  & 4.181    & 5.522      & 4.219    \\
\midrule
NP             & 38.321  & 3.646    & 3.562      & 3.586    \\
TP-OOV (b=0.4) & 46.813  & 3.846    & 3.103      & 3.916    \\
TP-OOV (b=1)   & 49.226  & 3.896    & 2.351      & 4.023  \\
\bottomrule
\end{tabular}
}
\end{table}

\begin{table}[t]
\centering
\caption{Compare TP-OOV with data filtering (OUT) on privacy and utility.
Privacy is evaluated with Loss-Ref MIA. 
N/A denotes a worse model performance than pre-training.
* denotes random guess in MIAs as $r=1$ means all samples are removed for OUT.
}
\label{tab:utility_out}
\resizebox{0.4\textwidth}{!}{
\begin{tabular}{c|c|ccccc}
\toprule
Metric                   & Method & r=0.05 & r=0.1 & r=0.5 & r=0.8  & r=1    \\
\midrule
\multirow{2}{*}{TPR@1\%} & OUT    & 0.000  & 0.557 & 0.047 & 0.004  & 0.01*  \\
                         & TP-OOV & 0.000  & 0.000 & 0.000 & 0.000  & 0.009  \\
\midrule
\multirow{2}{*}{AUC} & OUT    & 0.577  & 0.473 & 0.498 & 0.491  & 0.5*   \\
                         & TP-OOV & 0.552 
                         & 
                         0.463 	
                         & 0.504 & 0.494 & 0.494  \\
\midrule
\multirow{2}{*}{$\Delta$Val-PPL} & OUT    & 0.173  & 0.544 & 3.781 & 8.622  & 27.062 \\
                         & TP-OOV & 0.258  & 0.785 & 4.311 & 10.905 & N/A \\
\bottomrule
\end{tabular}
}
\end{table}

\noindent\textbf{Influences on Risk of Other Instances.}
We further investigate the broader influence of a self-guard on other instances, as a consideration for the privacy onion effect~\cite{carlini2022privacy}.
For the influence within defenders, the defense effectiveness in \Cref{tab:utility_out} is stable across various $r$.
This observation does not depend on the specific $b$ and methods as we leave similar results for the weakest variant UDP in Appendix.

Furthermore, we demonstrate the risk influence on the $1-r$ portion of unprotected samples in \Cref{fig:inf_r_un}.
When the majority of instances are self-guarded ($r=0.8$), the individual risk of unprotected instances is higher than with no protection, as unprotected samples become outliers when the outlier pattern has been rapidly learned by the model.
Considering the large amount of public web content, $r$ is usually small in practice.
In such cases, e.g., $r=0.05$, unprotected samples also benefit from a slightly lower risk than with no protection, attributed to the regularization effect introduced by perturbation~\cite{goodfellow2014explaining}.

\subsection{Robustness Analysis}\label{sec:robust}
As discussed in \Cref{sec:inv}, \ours is naturally robust to normal data pre-processing such as deduplication, and an active bypass requires hundreds of billions of verifications for perfect stripping.
Now we evaluate the robustness of \ours with two active strategies that an aggressive trainer takes: 1) performing active detection to locate self-guarded texts for manual stripping, and 2) conducting continuous training on clean data for recovering hidden protected text.\looseness=-1

\begin{figure*}[t]
    \centering
    \begin{subfigure}[b]{0.31\linewidth}
        \includegraphics[trim=0 0 0 20, clip, width=\linewidth]{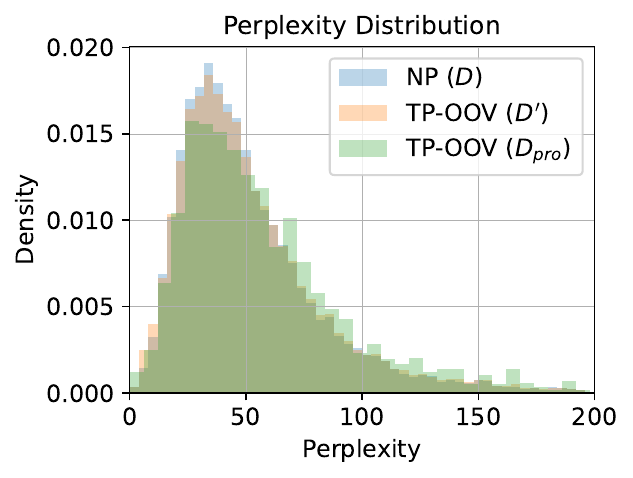}
        \caption{Perplexity distribution}
        \label{subfig:perplexity}
    \end{subfigure}
    \hfill
    \begin{subfigure}[b]{0.3\linewidth}
        \includegraphics[trim=0 0 0 20, clip, width=\linewidth]{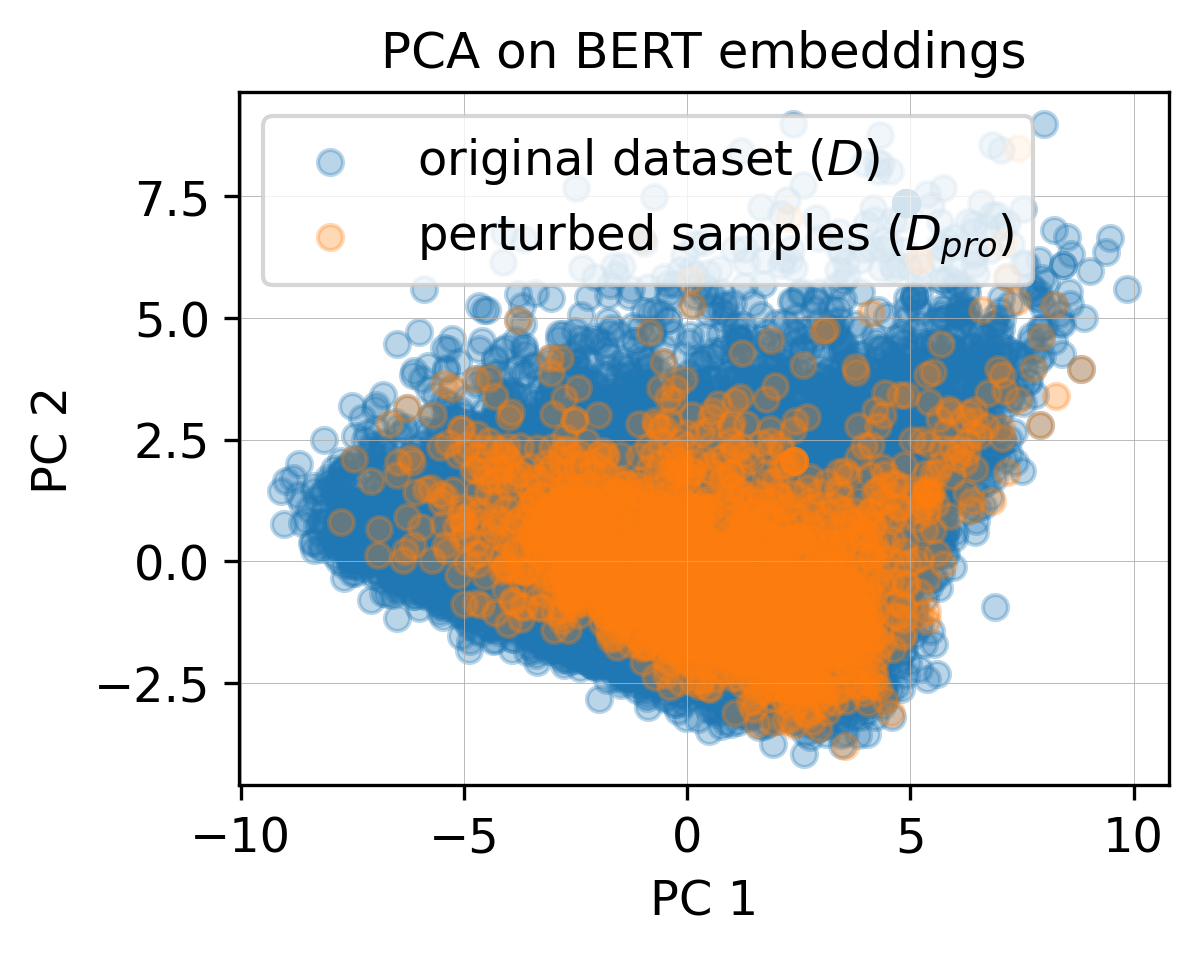}
        \caption{Embedding distribution (PCA)}
        \label{subfig:pca}
    \end{subfigure}
    \hfill
    \begin{subfigure}[b]{0.3\linewidth}
        \includegraphics[trim=0 7 0 5, clip, width=\linewidth]{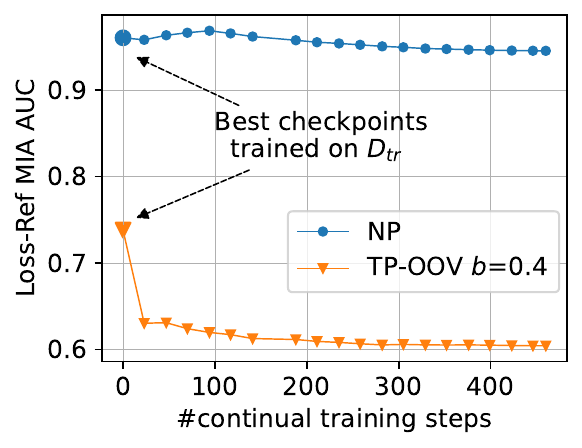}
        \caption{Effect of continual training}
        \label{subfig:continual}
    \end{subfigure}
    \caption{
    Robustness Analysis of Active Detection and Continuous Training.
    }
    \label{fig:robustness}
\end{figure*}

\noindent\textbf{Perturbation Filtering.}
Model trainers may use perplexity filtering~\cite{marion2023less} or embedding detection~\cite{kumar2020noisy} to improve data quality.
A large $b$ leads to poor-quality texts, so the whole sample can be filtered out.
In this case, there would be no violation on protected instances.
In the opposite case,  when $b$ is not high enough (such as $0.1$) to filter out the whole sample. \Cref{subfig:perplexity} shows that our proposed perturbation cannot be easily distinguished via both perplexity and embedding spaces. 
As a result, the perturbation remains in the protected text, acting as a shield.

\noindent\textbf{Effect of Continual Training with New Clean Data.} 
LLMs are commonly reused through continual training on new data~\cite{minaee2024}.
We evaluate robustness by fine-tuning GPT-2 on dataset $D$ (containing both protected and unprotected text), then continuing training on disjoint dataset $D_\text{new}$.
While prior work~\cite{chen2024janus} shows continual training can recover previously unexposed secrets, \Cref{subfig:continual} demonstrates that MIA risk on $D_\text{pro}$ steadily decreases as the model shifts focus to new data.
Notably, TP-OOV benefits more from continual training than NP, with AUC dropping from 0.75 to 0.6, confirming \ours's robustness to post-processing model updates.\looseness=-1

\section{Related Work}
As we aim to protect unauthorized content from being memorized and leaked by language models, our work is closely related to privacy defense and copyright protection.

\noindent\textbf{Privacy Defenses}.
There are substantial privacy defenses that can be applied to avoid generating training data from LMs, but all of them rely on collaboration from other parties.
In data pre-processing, deduplication~\cite{kandpal2022deduplicating} and scrubbing~\cite{lukas2023analyzing} require a trusted data curator, and it is hard to remove all sensitive information.
In the model training stage, differentially private (DP) optimization~\cite{martinabadiDeepLearningDifferential2016, liDifferentiallyPrivateMetaLearning2019, jayaramanEvaluatingDifferentiallyPrivate2019} ensures a theoretical privacy bound for each training record but requires a trusted model trainer who is willing to afford significantly higher training costs, especially for large LMs~\cite{liLargeLanguageModels2021,yuDifferentiallyPrivateFinetuning2021}.
Model alignment~\cite{hendrycks2020aligning} requires carefully designed alignment tasks.
In the inference stage, output filtering~\cite{GoyalAAAI24}, machine unlearning~\cite{caoMakingSystemsForget2015}, or model editing~\cite{yao2023editing} can be applied but require a trusted model curator and poses other risks~\cite{huang2024unlearn, hu2024learn}.
Distinguished from above privacy defenses, we aim to provide a broader protection, i.e., unauthorized content rather than only private content, and we do not rely on other parties.

\noindent\textbf{Copyright Protection}.
\label{sec:related_unlearn}
One strategy to protect copyrighted content is making it unlearnable.
The term unlearnable example~\cite{huang2021unlearnable, ren2022transferable, liu2024stable, fu2022robust, yu2022availability, li2023make} is proposed to prevent models from learning any knowledge from the perturbed dataset, and the success indicator is typically a poor inference-stage performance.
Unlike traditional unlearnable examples, we aim to degrade the exposure risk of an individual text.
Instead of image data or classification tasks in previous works, our defense targets language modeling tasks, requiring unique understanding of how generative models memorize training corpus.
The other strategy of copyright protection is to claim the data ownership by embedding data watermarks and detecting them via membership inference after model training~\cite{wei2024proving, zhaoprotecting}.
Although \ours was not designed for watermarking, we discuss how \ours can be used for data watermark in \Cref{sec:watermark-APP}.

\section{Discussion}\label{sec:discussion}
We discuss the benefits and limitations in practical scenarios.

\noindent\textbf{Easy-to-Use Protection.}
Besides the advantage of not relying on other parties, we naturally provide a personalized protection strength and scope.
The computational cost is affordable for defenders, even with proxy models, e.g., only taking <2 GB GPU memory for GPT-2 or millisecond-level API latency for online models.
Besides, the run-time overhead of perturbed webpage is small.
In our demonstration~\cite{expshield-demo-2025}, the loading time averaged over 10 trails only increase by $\approx$2\%.
The cost will be even lower for smaller budget (e.g., b=0.01 in \Cref{fig:eff_opt}).\looseness=-1

\noindent\textbf{Compatibility to Privacy Defenses.}
As an owner-side defense, \ours is compatible to other privacy defenses that may be implemented by other parties.
For example, since \ours has no repeated pattern and does not contain meaningful entities, it is robust when integrating with privacy defenses in preprocessing, such as deduplication~\cite{kandpal2022deduplicating} and scrubbing~\cite{lukas2023analyzing}.
Also, the extra layer of \ours does not violate the theoretical guarantee of DP  training~\cite{abadi2016deep} and other inference phase defenses~\cite{panaitescu2024can}.\looseness=-1

\noindent\textbf{Generalization to Other Languages.}
In principle, \ours~can be used for other languages, as it operates at the token level without relying on English-specific syntax, independent of specific word segmentation.
Specifically, our OOV-based perturbation also applies to CJK characters, because several CJK characters can be merged into a single token in popular tokenizers (BPE, SentencePiece).
For languages such as Thai or Khmar where zero-width character is used for line breaking, we suggest to use style-level perturbation to avoid rendering issues.\looseness=-1

\noindent\textbf{Against High-Resource Crawlers.}
While our main target is large-scale automatic crawlers instead of aggressive crawlers with high-resource computation, \ours is still possible to reduce risks because there is no guarantee for perfect sanitization.
In \Cref{fig:eff_opt}, only $\approx1\%$ surviving perturbation still lower memorization risk.
Defenders can even combine multiple \ours variants to enhance the complexity of sanitization.
While Optical Character Recognition (OCR) is powerful, mainstream LLMs~\cite{dubey2024llama,gao2020pile800gbdatasetdiverse} use HTML extraction instead of OCR for collecting large-scale webpages, probably due to its accuracy and cost (around 100–300× more costly).\looseness=-1

\noindent\textbf{Impacts on User Experience.}
The negative impacts on the user experience of perturbed webpages are minimal.
Website owners can exclude the self-injected perturbation to ensure accurate internal site search, and mark perturbed pages as \texttt{noindex} to avoid the mismatch issue on legitimate search engine optimization (SEO).
In the extreme case where every word is perturbed by invisible characters, browser search, word selection, screen reading, text highlighting are unaffected; the translation remains functional for most words. And the style-level perturbation is functional with a small budget (e.g., $b<0.1$).
In \Cref{fig:eff_opt}, even a small budget demonstrates the effectiveness of the defense, so the overall impact is marginal.
\looseness=-1

\noindent\textbf{Collaborative Mitigation of Data Misuse.}
When protected data is unintentionally misused, such as being collected by third-party crawlers, model trainers who are aware of the defense are encouraged to either:
(i) exclude all segments that may contain perturbations to avoid degrading training quality and data misuse; or
(ii) include the perturbed data during training while filtering out invisible characters at inference time, both to reduce the risk of privacy attack and mitigate safety-related abuses.\looseness=-1

\section{Conclusion}
We present \ours, a proactive self-guard that empowers data owners with direct control over their content's usage in AI training, addressing ineffective crawl prevention and third-party dependency.

Our approach fills the critical gap where content creators lack protection against unauthorized LLM memorization through a practical, independent solution requiring no third-party cooperation.
We formalize individual text protection as constrained bi-level optimization that minimizes adversarial advantage while preserving readability and budget constraints.
For principled evaluation and design, we introduce instance exploitation—a standard, calibrated, and efficient individual-level privacy metric that is informative for a wide scope of attacks.
By establishing and verifying the memorization trigger hypothesis, we develop targeted perturbations that focus on influencing important tokens for memorization.
It is promising for future works to design more informed individual-risk metrics and extend the memorization trigger hypothesis for improving defenses during or after training. \looseness=-1

We comprehensively validate the defense effectiveness across various tasks with LMs and VLMs, showing its capability to reduce the near-perfect attack to random membership guess.
Additionally, we extend discussions of its feasibility in practice and advocate a collaborative view for responsible AI where data protection rights and innovation coexist.

\section*{Acknowledgment}
We would like to thank reviewers for their constructive comments and efforts in improving our paper.
This work was supported in part by  
National Science Foundation grants (CNS-2437345, CNS-2125530, CNS-2124104, IIS-2302968, CNS-2350333) and 
National Institutes of Health grants (R01LM013712, R01ES033241).

\bibliographystyle{IEEEtran}
\bibliography{bib/crypto,bib/llm,bib/ppml,bib/security,bib/toan,bib/unlearn}

\begin{thebibliography}{10}
\providecommand{\url}[1]{#1}
\csname url@samestyle\endcsname
\providecommand{\newblock}{\relax}
\providecommand{\bibinfo}[2]{#2}
\providecommand{\BIBentrySTDinterwordspacing}{\spaceskip=0pt\relax}
\providecommand{\BIBentryALTinterwordstretchfactor}{4}
\providecommand{\BIBentryALTinterwordspacing}{\spaceskip=\fontdimen2\font plus
\BIBentryALTinterwordstretchfactor\fontdimen3\font minus \fontdimen4\font\relax}
\providecommand{\BIBforeignlanguage}[2]{{%
\expandafter\ifx\csname l@#1\endcsname\relax
\typeout{** WARNING: IEEEtran.bst: No hyphenation pattern has been}%
\typeout{** loaded for the language `#1'. Using the pattern for}%
\typeout{** the default language instead.}%
\else
\language=\csname l@#1\endcsname
\fi
#2}}
\providecommand{\BIBdecl}{\relax}
\BIBdecl

\bibitem{brown2020language}
T.~Brown, B.~Mann, N.~Ryder, M.~Subbiah, J.~D. Kaplan, P.~Dhariwal, A.~Neelakantan, P.~Shyam, G.~Sastry, A.~Askell \emph{et~al.}, ``Language models are few-shot learners,'' \emph{Advances in neural information processing systems}, vol.~33, pp. 1877--1901, 2020.

\bibitem{henderson2023foundation}
P.~Henderson, X.~Li, D.~Jurafsky, T.~Hashimoto, M.~A. Lemley, and P.~Liang, ``Foundation models and fair use,'' \emph{Journal of Machine Learning Research}, vol.~24, no. 400, pp. 1--79, 2023.

\bibitem{tramer2024position}
F.~Tram{\`e}r, G.~Kamath, and N.~Carlini, ``Position: Considerations for differentially private learning with large-scale public pretraining,'' in \emph{Forty-first International Conference on Machine Learning}, 2024.

\bibitem{carlini2021extracting}
N.~Carlini, F.~Tramer, E.~Wallace, M.~Jagielski, A.~Herbert-Voss, K.~Lee, A.~Roberts, T.~Brown, D.~Song, U.~Erlingsson \emph{et~al.}, ``Extracting training data from large language models,'' in \emph{30th USENIX Security Symposium (USENIX Security 21)}, 2021, pp. 2633--2650.

\bibitem{freeman2024exploring}
J.~Freeman, C.~Rippe, E.~Debenedetti, and M.~Andriushchenko, ``Exploring memorization and copyright violation in frontier llms: A study of the new york times v. openai 2023 lawsuit,'' \emph{arXiv preprint arXiv:2412.06370}, 2024.

\bibitem{loh2021social}
J.~Loh and M.~J. Walsh, ``Social media context collapse: The consequential differences between context collusion versus context collision,'' \emph{Social Media+ Society}, vol.~7, no.~3, p. 20563051211041646, 2021.

\bibitem{khder2021web}
M.~A. Khder, ``Web scraping or web crawling: State of art, techniques, approaches and application.'' \emph{International Journal of Advances in Soft Computing \& Its Applications}, vol.~13, no.~3, 2021.

\bibitem{kandpal2022deduplicating}
N.~Kandpal, E.~Wallace, and C.~Raffel, ``Deduplicating training data mitigates privacy risks in language models,'' in \emph{International Conference on Machine Learning}.\hskip 1em plus 0.5em minus 0.4em\relax PMLR, 2022, pp. 10\,697--10\,707.

\bibitem{lukas2023analyzing}
N.~Lukas, A.~Salem, R.~Sim, S.~Tople, L.~Wutschitz, and S.~Zanella-B{\'e}guelin, ``Analyzing leakage of personally identifiable information in language models,'' in \emph{2023 IEEE Symposium on Security and Privacy (SP)}.\hskip 1em plus 0.5em minus 0.4em\relax IEEE, 2023, pp. 346--363.

\bibitem{abadi2016deep}
M.~Abadi, A.~Chu, I.~Goodfellow, H.~B. McMahan, I.~Mironov, K.~Talwar, and L.~Zhang, ``Deep learning with differential privacy,'' in \emph{Proceedings of the 2016 ACM SIGSAC Conference on Computer and Communications Security}, 2016, pp. 308--318.

\bibitem{hendrycks2020aligning}
\BIBentryALTinterwordspacing
D.~Hendrycks, C.~Burns, S.~Basart, A.~Critch, J.~Li, D.~Song, and J.~Steinhardt, ``Aligning {\{}ai{\}} with shared human values,'' in \emph{International Conference on Learning Representations}, 2021. [Online]. Available: \url{https://openreview.net/forum?id=dNy_RKzJacY}
\BIBentrySTDinterwordspacing

\bibitem{panaitescu2024can}
\BIBentryALTinterwordspacing
M.-A. Panaitescu-Liess, Z.~Che, B.~An, Y.~Xu, P.~Pathmanathan, S.~Chakraborty, S.~Zhu, T.~Goldstein, and F.~Huang, ``Can watermarking large language models prevent copyrighted text generation and hide training data?'' in \emph{Proceedings of the Thirty-Ninth AAAI Conference on Artificial Intelligence and Thirty-Seventh Conference on Innovative Applications of Artificial Intelligence and Fifteenth Symposium on Educational Advances in Artificial Intelligence}, ser. AAAI'25/IAAI'25/EAAI'25.\hskip 1em plus 0.5em minus 0.4em\relax AAAI Press, 2025. [Online]. Available: \url{https://doi.org/10.1609/aaai.v39i23.34684}
\BIBentrySTDinterwordspacing

\bibitem{igamberdiev2023dp}
\BIBentryALTinterwordspacing
T.~Igamberdiev and I.~Habernal, ``{DP}-{BART} for privatized text rewriting under local differential privacy,'' in \emph{Findings of the Association for Computational Linguistics: ACL 2023}, A.~Rogers, J.~Boyd-Graber, and N.~Okazaki, Eds.\hskip 1em plus 0.5em minus 0.4em\relax Toronto, Canada: Association for Computational Linguistics, Jul. 2023, pp. 13\,914--13\,934. [Online]. Available: \url{https://aclanthology.org/2023.findings-acl.874/}
\BIBentrySTDinterwordspacing

\bibitem{yue2021differential}
\BIBentryALTinterwordspacing
X.~Yue, M.~Du, T.~Wang, Y.~Li, H.~Sun, and S.~S.~M. Chow, ``Differential privacy for text analytics via natural text sanitization,'' in \emph{Findings of the Association for Computational Linguistics: ACL-IJCNLP 2021}.\hskip 1em plus 0.5em minus 0.4em\relax Online: Association for Computational Linguistics, Aug. 2021, pp. 3853--3866. [Online]. Available: \url{https://aclanthology.org/2021.findings-acl.337/}
\BIBentrySTDinterwordspacing

\bibitem{li2023make}
\BIBentryALTinterwordspacing
X.~Li and M.~Liu, ``Make text unlearnable: Exploiting effective patterns to protect personal data,'' in \emph{Proceedings of the 3rd Workshop on Trustworthy Natural Language Processing (TrustNLP 2023)}.\hskip 1em plus 0.5em minus 0.4em\relax Association for Computational Linguistics, Jul. 2023, pp. 249--259. [Online]. Available: \url{https://aclanthology.org/2023.trustnlp-1.22/}
\BIBentrySTDinterwordspacing

\bibitem{carlini2022membership}
N.~Carlini, S.~Chien, M.~Nasr, S.~Song, A.~Terzis, and F.~Tramer, ``Membership inference attacks from first principles,'' in \emph{2022 IEEE Symposium on Security and Privacy (SP)}.\hskip 1em plus 0.5em minus 0.4em\relax IEEE, 2022, pp. 1897--1914.

\bibitem{carliniSecretSharerEvaluating2019}
N.~Carlini, C.~Liu, {\'U}.~Erlingsson, J.~Kos, and D.~Song, ``The {{Secret Sharer}}: {{Evaluating}} and {{Testing Unintended Memorization}} in {{Neural Networks}},'' \emph{arXiv:1802.08232 [cs]}, Jul. 2019.

\bibitem{fowl2021adversarial}
L.~Fowl, M.~Goldblum, P.-y. Chiang, J.~Geiping, W.~Czaja, and T.~Goldstein, ``Adversarial examples make strong poisons,'' \emph{Advances in Neural Information Processing Systems}, 2021.

\bibitem{yu2022availability}
D.~Yu, H.~Zhang, W.~Chen, J.~Yin, and T.-Y. Liu, ``Availability attacks create shortcuts,'' in \emph{Proceedings of the 28th ACM SIGKDD Conference on Knowledge Discovery and Data Mining}, 2022, pp. 2367--2376.

\bibitem{huang2021unlearnable}
H.~Huang, X.~Ma, S.~M. Erfani, J.~Bailey, and Y.~Wang, ``Unlearnable examples: Making personal data unexploitable,'' \emph{arXiv preprint arXiv:2101.04898}, 2021.

\bibitem{bowen2024}
\BIBentryALTinterwordspacing
D.~Bowen, B.~Murphy, W.~Cai, D.~Khachaturov, A.~Gleave, and K.~Pelrine, ``Data poisoning in llms: Jailbreak-tuning and scaling laws,'' 2024. [Online]. Available: \url{https://arxiv.org/abs/2408.02946}
\BIBentrySTDinterwordspacing

\bibitem{liu2024precurious}
R.~Liu, T.~Wang, Y.~Cao, and L.~Xiong, ``Precurious: How innocent pre-trained language models turn into privacy traps,'' in \emph{Proceedings of the 2024 ACM SIGSAC Conference on Computer and Communications Security}, 2024.

\bibitem{wen2024privacy}
Y.~Wen, L.~Marchyok, S.~Hong, J.~Geiping, T.~Goldstein, and N.~Carlini, ``Privacy backdoors: Enhancing membership inference through poisoning pre-trained models,'' \emph{arXiv preprint arXiv:2404.01231}, 2024.

\bibitem{nasr2023scalable}
M.~Nasr, N.~Carlini, J.~Hayase, M.~Jagielski, A.~F. Cooper, D.~Ippolito, C.~A. Choquette-Choo, E.~Wallace, F.~Tram{\`e}r, and K.~Lee, ``Scalable extraction of training data from (production) language models,'' \emph{arXiv preprint arXiv:2311.17035}, 2023.

\bibitem{vaswani2017attention}
A.~Vaswani, ``Attention is all you need,'' \emph{Advances in Neural Information Processing Systems}, 2017.

\bibitem{radford2018improving}
A.~Radford, ``Improving language understanding by generative pre-training,'' 2018.

\bibitem{hayes2024measuring}
J.~Hayes, M.~Swanberg, H.~Chaudhari, I.~Yona, I.~Shumailov, M.~Nasr, C.~A. Choquette-Choo, K.~Lee, and A.~F. Cooper, ``Measuring memorization in language models via probabilistic extraction,'' \emph{arXiv preprint arXiv:2410.19482}, 2024.

\bibitem{yeomPrivacyRiskMachine2018}
S.~Yeom, I.~Giacomelli, M.~Fredrikson, and S.~Jha, ``Privacy {{Risk}} in {{Machine Learning}}: {{Analyzing}} the {{Connection}} to {{Overfitting}},'' \emph{arXiv:1709.01604 [cs, stat]}, May 2018.

\bibitem{dworkDifferentialPrivacyPractice2019}
C.~Dwork, N.~Kohli, and D.~Mulligan, ``Differential {{Privacy}} in {{Practice}}: {{Expose}} your {{Epsilons}}!'' \emph{Journal of Privacy and Confidentiality}, vol.~9, no.~2, Oct. 2019.

\bibitem{balle2022reconstructing}
B.~Balle, G.~Cherubin, and J.~Hayes, ``Reconstructing training data with informed adversaries,'' in \emph{2022 IEEE Symposium on Security and Privacy (SP)}.\hskip 1em plus 0.5em minus 0.4em\relax IEEE, 2022, pp. 1138--1156.

\bibitem{salem2023sok}
A.~Salem, G.~Cherubin, D.~Evans, B.~K{\"o}pf, A.~Paverd, A.~Suri, S.~Tople, and S.~Zanella-B{\'e}guelin, ``Sok: Let the privacy games begin! a unified treatment of data inference privacy in machine learning,'' in \emph{2023 IEEE Symposium on Security and Privacy (SP)}.\hskip 1em plus 0.5em minus 0.4em\relax IEEE, 2023, pp. 327--345.

\bibitem{massey1951kolmogorov}
F.~J. Massey~Jr, ``The kolmogorov-smirnov test for goodness of fit,'' \emph{Journal of the American statistical Association}, vol.~46, no. 253, pp. 68--78, 1951.

\bibitem{carlini2022privacy}
N.~Carlini, M.~Jagielski, C.~Zhang, N.~Papernot, A.~Terzis, and F.~Tramer, ``The privacy onion effect: Memorization is relative,'' \emph{Advances in Neural Information Processing Systems}, vol.~35, pp. 13\,263--13\,276, 2022.

\bibitem{boucher2022bad}
N.~Boucher, I.~Shumailov, R.~Anderson, and N.~Papernot, ``Bad characters: Imperceptible nlp attacks,'' in \emph{2022 IEEE Symposium on Security and Privacy (SP)}.\hskip 1em plus 0.5em minus 0.4em\relax IEEE, 2022, pp. 1987--2004.

\bibitem{liao2024eia}
Z.~Liao, L.~Mo, C.~Xu, M.~Kang, J.~Zhang, C.~Xiao, Y.~Tian, B.~Li, and H.~Sun, ``Eia: Environmental injection attack on generalist web agents for privacy leakage,'' \emph{arXiv preprint arXiv:2409.11295}, 2024.

\bibitem{richardson2007beautiful}
L.~Richardson, ``Beautiful soup documentation,'' 2007.

\bibitem{expshield-demo-2025}
{ExpShield's authors}, ``Expshield demo,'' \url{https://github.com/toan-vt/ExpShield-demo}, 2025, gitHub-style repository, accessed 7 Aug 2025.

\bibitem{broder1997resemblance}
A.~Z. Broder, ``On the resemblance and containment of documents,'' in \emph{Proceedings. Compression and Complexity of SEQUENCES 1997 (Cat. No. 97TB100171)}.\hskip 1em plus 0.5em minus 0.4em\relax IEEE, 1997, pp. 21--29.

\bibitem{longpre2024pretrainer}
S.~Longpre, G.~Yauney, E.~Reif, K.~Lee, A.~Roberts, B.~Zoph, D.~Zhou, J.~Wei, K.~Robinson, D.~Mimno \emph{et~al.}, ``A pretrainer’s guide to training data: Measuring the effects of data age, domain coverage, quality, \& toxicity,'' in \emph{Proceedings of the 2024 Conference of the North American Chapter of the Association for Computational Linguistics: Human Language Technologies (Volume 1: Long Papers)}, 2024, pp. 3245--3276.

\bibitem{anonymous2025do}
\BIBentryALTinterwordspacing
Anonymous, ``Do we really have to filter out random noise in pre-training data for language models?'' in \emph{Submitted to ACL Rolling Review - February 2025}, 2025, under review. [Online]. Available: \url{https://openreview.net/forum?id=mpnPR8YQ3d}
\BIBentrySTDinterwordspacing

\bibitem{shi2023detecting}
W.~Shi, A.~Ajith, M.~Xia, Y.~Huang, D.~Liu, T.~Blevins, D.~Chen, and L.~Zettlemoyer, ``Detecting pretraining data from large language models,'' \emph{arXiv preprint arXiv:2310.16789}, 2023.

\bibitem{zou2023universal}
A.~Zou, Z.~Wang, N.~Carlini, M.~Nasr, J.~Z. Kolter, and M.~Fredrikson, ``Universal and transferable adversarial attacks on aligned language models,'' \emph{arXiv preprint arXiv:2307.15043}, 2023.

\bibitem{klimt2004enron}
B.~Klimt and Y.~Yang, ``The enron corpus: A new dataset for email classification research,'' in \emph{European Conference on Machine Learning}.\hskip 1em plus 0.5em minus 0.4em\relax Springer, 2004, pp. 217--226.

\bibitem{zeng2020meddialog}
G.~Zeng, W.~Yang, Z.~Ju, Y.~Yang, S.~Wang, R.~Zhang, M.~Zhou, J.~Zeng, X.~Dong, R.~Zhang \emph{et~al.}, ``Meddialog: Large-scale medical dialogue datasets,'' in \emph{Proceedings of the 2020 Conference on Empirical Methods in Natural Language Processing (EMNLP)}, 2020.

\bibitem{mackenzie2020cc}
J.~Mackenzie, R.~Benham, M.~Petri, J.~R. Trippas, J.~S. Culpepper, and A.~Moffat, ``Cc-news-en: A large english news corpus,'' in \emph{Proceedings of the 29th ACM International Conference on Information \& Knowledge Management}, 2020, pp. 3077--3084.

\bibitem{iapr}
M.~Grubinger, P.~Clough, H.~Müller, and T.~Deselaers, ``The iapr tc12 benchmark: A new evaluation resource for visual information systems,'' \emph{International Conference on Language Resources and Evaluation}, 2006.

\bibitem{radford2019language}
A.~Radford, J.~Wu, R.~Child, D.~Luan, D.~Amodei, I.~Sutskever \emph{et~al.}, ``Language models are unsupervised multitask learners,'' \emph{OpenAI blog}, vol.~1, no.~8, p.~9, 2019.

\bibitem{zhang2022opt}
S.~Zhang, S.~Roller, N.~Goyal, M.~Artetxe, M.~Chen, S.~Chen, C.~Dewan, M.~Diab, X.~Li, X.~V. Lin \emph{et~al.}, ``Opt: Open pre-trained transformer language models,'' \emph{arXiv preprint arXiv:2205.01068}, 2022.

\bibitem{touvron2023llama}
H.~Touvron, L.~Martin, K.~Stone, P.~Albert, A.~Almahairi, Y.~Babaei, N.~Bashlykov, S.~Batra, P.~Bhargava, S.~Bhosale \emph{et~al.}, ``Llama 2: Open foundation and fine-tuned chat models,'' \emph{arXiv preprint arXiv:2307.09288}, 2023.

\bibitem{efron1992bootstrap}
B.~Efron, ``Bootstrap methods: another look at the jackknife,'' in \emph{Breakthroughs in Statistics: Methodology and Distribution}.\hskip 1em plus 0.5em minus 0.4em\relax Springer, 1992, pp. 569--593.

\bibitem{puerto2024scaling}
H.~Puerto, M.~Gubri, S.~Yun, and S.~J. Oh, ``Scaling up membership inference: When and how attacks succeed on large language models,'' \emph{arXiv preprint arXiv:2411.00154}, 2024.

\bibitem{goodfellow2014explaining}
I.~J. Goodfellow, J.~Shlens, and C.~Szegedy, ``Explaining and harnessing adversarial examples,'' \emph{arXiv preprint arXiv:1412.6572}, 2014.

\bibitem{marion2023less}
M.~Marion, A.~{\"U}st{\"u}n, L.~Pozzobon, A.~Wang, M.~Fadaee, and S.~Hooker, ``When less is more: Investigating data pruning for pretraining llms at scale,'' \emph{arXiv preprint arXiv:2309.04564}, 2023.

\bibitem{kumar2020noisy}
A.~Kumar, P.~Makhija, and A.~Gupta, ``Noisy text data: Achilles' heel of bert,'' \emph{arXiv preprint arXiv:2003.12932}, 2020.

\bibitem{minaee2024}
\BIBentryALTinterwordspacing
S.~Minaee, T.~Mikolov, N.~Nikzad, M.~Chenaghlu, R.~Socher, X.~Amatriain, and J.~Gao, ``Large language models: A survey,'' 2024. [Online]. Available: \url{https://arxiv.org/abs/2402.06196}
\BIBentrySTDinterwordspacing

\bibitem{chen2024janus}
X.~Chen, S.~Tang, R.~Zhu, S.~Yan, L.~Jin, Z.~Wang, L.~Su, Z.~Zhang, X.~Wang, and H.~Tang, ``The janus interface: How fine-tuning in large language models amplifies the privacy risks,'' in \emph{Proceedings of the 2024 on ACM SIGSAC Conference on Computer and Communications Security}, 2024, pp. 1285--1299.

\bibitem{martinabadiDeepLearningDifferential2016}
{Mart{\'i}n Abadi}, A.~Chu, I.~Goodfellow, H.~B. McMahan, I.~Mironov, K.~Talwar, and L.~Zhang, ``Deep {{Learning}} with {{Differential Privacy}},'' \emph{Proceedings of the 2016 ACM SIGSAC Conference on Computer and Communications Security - CCS'16}, pp. 308--318, 2016.

\bibitem{liDifferentiallyPrivateMetaLearning2019}
J.~Li, M.~Khodak, S.~Caldas, and A.~Talwalkar, ``Differentially {{Private Meta-Learning}},'' \emph{arXiv:1909.05830 [cs, stat]}, Sep. 2019.

\bibitem{jayaramanEvaluatingDifferentiallyPrivate2019}
B.~Jayaraman and D.~Evans, ``Evaluating {{Differentially Private Machine Learning}} in {{Practice}},'' in \emph{{{USENIX}}}, 2019, p.~18.

\bibitem{liLargeLanguageModels2021}
X.~Li, F.~Tram{\`e}r, P.~Liang, and T.~Hashimoto, ``Large {{Language Models Can Be Strong Differentially Private Learners}},'' \emph{arXiv:2110.05679 [cs]}, Oct. 2021.

\bibitem{yuDifferentiallyPrivateFinetuning2021}
D.~Yu, S.~Naik, A.~Backurs, S.~Gopi, H.~A. Inan, G.~Kamath, J.~Kulkarni, Y.~T. Lee, A.~Manoel, L.~Wutschitz, S.~Yekhanin, and H.~Zhang, ``Differentially {{Private Fine-tuning}} of {{Language Models}},'' \emph{arXiv:2110.06500 [cs, stat]}, Oct. 2021.

\bibitem{GoyalAAAI24}
S.~Goyal, M.~Hira, S.~Mishra, S.~Goyal, A.~Goel, N.~Dadu, K.~DB, S.~Mehta, and N.~Madaan, ``Llmguard: Guarding against unsafe llm behavior,'' \emph{Proceedings of the AAAI Conference on Artificial Intelligence}, vol.~38, pp. 23\,790--23\,792, Mar. 2024.

\bibitem{caoMakingSystemsForget2015}
Y.~Cao and J.~Yang, ``Towards {{Making Systems Forget}} with {{Machine Unlearning}},'' in \emph{2015 IEEE Symposium on Security and Privacy (SP)}.\hskip 1em plus 0.5em minus 0.4em\relax San Jose, CA: IEEE, May 2015, pp. 463--480.

\bibitem{yao2023editing}
Y.~Yao, P.~Wang, B.~Tian, S.~Cheng, Z.~Li, S.~Deng, H.~Chen, and N.~Zhang, ``Editing large language models: Problems, methods, and opportunities,'' in \emph{Proceedings of the 2023 Conference on Empirical Methods in Natural Language Processing}, 2023.

\bibitem{huang2024unlearn}
Y.~Huang, D.~Liu, L.~Chua, B.~Ghazi, P.~Kamath, R.~Kumar, P.~Manurangsi, M.~Nasr, A.~Sinha, and C.~Zhang, ``Unlearn and burn: Adversarial machine unlearning requests destroy model accuracy,'' \emph{arXiv preprint arXiv:2410.09591}, 2024.

\bibitem{hu2024learn}
H.~Hu, S.~Wang, T.~Dong, and M.~Xue, ``Learn what you want to unlearn: Unlearning inversion attacks against machine unlearning,'' in \emph{2024 IEEE Symposium on Security and Privacy (SP)}.\hskip 1em plus 0.5em minus 0.4em\relax IEEE Computer Society, 2024, pp. 262--262.

\bibitem{ren2022transferable}
J.~Ren, H.~Xu, Y.~Wan, X.~Ma, L.~Sun, and J.~Tang, ``Transferable unlearnable examples,'' \emph{arXiv preprint arXiv:2210.10114}, 2022.

\bibitem{liu2024stable}
Y.~Liu, K.~Xu, X.~Chen, and L.~Sun, ``Stable unlearnable example: Enhancing the robustness of unlearnable examples via stable error-minimizing noise,'' in \emph{Proceedings of the AAAI Conference on Artificial Intelligence}, vol.~38, no.~4, 2024, pp. 3783--3791.

\bibitem{fu2022robust}
S.~Fu, F.~He, Y.~Liu, L.~Shen, and D.~Tao, ``Robust unlearnable examples: Protecting data against adversarial learning,'' \emph{arXiv preprint arXiv:2203.14533}, 2022.

\bibitem{wei2024proving}
J.~T.-Z. Wei, R.~Y. Wang, and R.~Jia, ``Proving membership in llm pretraining data via data watermarks,'' \emph{arXiv preprint arXiv:2402.10892}, 2024.

\bibitem{zhaoprotecting}
\BIBentryALTinterwordspacing
S.~Zhao, L.~Zhu, R.~Quan, and Y.~Yang, ``Protecting copyrighted material with unique identifiers in large language model training,'' 2025. [Online]. Available: \url{https://arxiv.org/abs/2403.15740}
\BIBentrySTDinterwordspacing

\bibitem{dubey2024llama}
A.~Dubey, A.~Jauhri, A.~Pandey, A.~Kadian, A.~Al-Dahle, A.~Letman, A.~Mathur, A.~Schelten, A.~Yang, A.~Fan \emph{et~al.}, ``The llama 3 herd of models,'' \emph{arXiv e-prints}, pp. arXiv--2407, 2024.

\bibitem{gao2020pile800gbdatasetdiverse}
\BIBentryALTinterwordspacing
L.~Gao, S.~Biderman, S.~Black, L.~Golding, T.~Hoppe, C.~Foster, J.~Phang, H.~He, A.~Thite, N.~Nabeshima, S.~Presser, and C.~Leahy, ``The pile: An 800gb dataset of diverse text for language modeling,'' 2020. [Online]. Available: \url{https://arxiv.org/abs/2101.00027}
\BIBentrySTDinterwordspacing

\bibitem{blip2}
J.~Li, D.~Li, S.~Savarese, and S.~Hoi, ``Blip-2: bootstrapping language-image pre-training with frozen image encoders and large language models,'' in \emph{Proceedings of the 40th International Conference on Machine Learning}, ser. ICML'23.\hskip 1em plus 0.5em minus 0.4em\relax JMLR.org, 2023.

\bibitem{dittrich2012menlo}
D.~Dittrich, E.~Kenneally \emph{et~al.}, ``The menlo report: Ethical principles guiding information and communication technology research,'' US Department of Homeland Security, Tech. Rep., 2012.

\end{thebibliography}

\appendix

\subsection{Dataset Information}\label{sec:data-APP}
\begin{itemize}
\item Enron~\cite{klimt2004enron} is a large collection of email data from the Enron Corporation, which contains sufficient PII information such as phone numbers, email addresses, and names. The dataset comprises emails from 150 users, primarily senior management of Enron.
\item Patient~\cite{zeng2020meddialog} consists of doctor-patient conversations covering various medical conditions, symptoms, diagnoses, and treatment plans, with an average length of 8 turns per conversation.
\item CC-News~\cite{mackenzie2020cc} is derived from the Common Crawl News dataset, containing news articles from various online sources published between 2016-2019. The articles span diverse topics and writing styles, providing a rich test bed for evaluating privacy preservation in copyrighted content.
\end{itemize}

\begin{table}[htbp]
\centering
\caption{Defense Strategy Comparison on Scenarios and Compatibility}
\label{tab:defense-matrix_APP}
\resizebox{0.5\textwidth}{!}{
\begin{tabular}{@{}lllll@{}}
\toprule
\textbf{Defense Type} & 
\begin{tabular}{@{}l@{}}\textbf{Personal}\\ \textbf{Websites}\end{tabular} & 
\begin{tabular}{@{}l@{}}\textbf{UGC}\\ \textbf{Platforms}\end{tabular} & 
\begin{tabular}{@{}l@{}}\textbf{Screen Reader}\\ \textbf{Compatibility}\end{tabular} & 
\begin{tabular}{@{}l@{}}\textbf{DOM Element Count}\\ \textbf{Impact}\end{tabular} \\
\midrule

Style-Level & 
\checkmark & 
\texttimes & 
Medium  
& 
Slightly increase 
\\

\addlinespace[0.5em]

Character-Level & 
\checkmark & 
\checkmark & 
High 
& 
\begin{tabular}{@{}l@{}}
No change
\end{tabular} \\

\bottomrule
\end{tabular}
}
\end{table}

\section{Perturbation Implementations}\label{sec:app_perturb_imp}
\subsection{Perturbation operation discussions}

A previous work~\cite{boucher2022bad} proposes adversarial examples by inserting imperceptible characters, such as invisible characters, homoglyphs, reordering characters, and deletion characters, into text inputs. These perturbations as follows are designed to be undetectable by human users while significantly altering the output of natural language processing (NLP) models.

\begin{itemize}
    \item \textbf{Augmentation}: Augment original text by modifying the encoding style without altering its visual display. 
Techniques include applying CSS properties like font-size or absolute position to make text invisible, inserting zero-width or invisible whitespace characters, and hiding text within HTML comment tags.
In our TP-OOV, examples of invisible characters include the Zero Width Space (U+200B), Zero Width Non-Joiner (U+200C), and Zero Width Joiner (U+200D). These characters do not render visually but are encoded in the HTML, allowing for subtle modifications. 
    
    \item \textbf{Deletion}: Remove characters to obscure text, either through delete-characters (e.g., Backspace, Delete) that are font and platform-independent, or using JavaScript to conditionally hide content, making it platform-dependent. 
These approaches are effective against basic scrapers but less so against those that can execute JavaScript.
CSS pseudo-elements such as `::before' or `::after' and replacing hidden text with SVG graphics can also hide partial text.

    \item \textbf{Replacement}: Replace characters with HTML entities, or visually similar homoglyphs (e.g., replacing Latin letters with visually similar Cyrillic ones), which are character-dependent. This technique confuses basic scrapers but may lead to imperfect readability.

    \item \textbf{Shuffling}: Use control characters like Carriage Return (CR), Backspace (BS), or Delete (DEL) to reorder or hide parts of the text. This method is platform- and character-dependent, effective at shuffling content without reducing readability when done carefully.
\end{itemize}

Furthermore, we compare character-level and style-level perturbation with respect to practical application in \Cref{tab:defense-matrix_APP}.

\subsection{Demonstration of invisible perturbation}\label{APP:sec:invisible}

\begin{figure}[htbp]
\begin{tcolorbox}[
  title=Three Methods to Make Text Invisible in HTML,
  colback=background,
  colframe=blue!75!black,
  coltitle=white,
  fonttitle=\bfseries,
  boxrule=0.5mm
]
\begin{lstlisting}[style=HTMLstyle]
<!-- Method 1: Using CSS display property -->
<div style="display: none;">This text is invisible</div>

<!-- Method 2: Using zero-width characters -->
<span>Visible text&#8203;z&#8203;e&#8203;r&#8203;o&#8203;-&#8203;w&#8203;i&#8203;d&#8203;t&#8203;h&#8203; characters hidden here</span>

<!-- Method 3: Using CSS positioning and size -->
<div style="position: absolute; left: -9999px; font-size: 0;">
  This text is positioned off-screen and has zero font size
</div>
\end{lstlisting}
\end{tcolorbox}
\caption{A simplified demonstration of invisibility strategy}
\label{fig:html_demo}
\end{figure}

We demonstrate a few ways of creating invisible styles in \Cref{fig:html_demo} and provide a concrete example in demonstration~\cite{expshield-demo-2025}:
\begin{itemize}
  \item \textbf{Method 1:} Uses CSS \texttt{display: none} to prevent the element from rendering in the document flow.
  \item \textbf{Method 2:} Inserts zero-width space characters (Unicode U+200B) between letters, making the text invisible while maintaining its position in the document.
  \item \textbf{Method 3:} Combines absolute positioning (moving the element far off-screen) with zero font size to hide text.
\end{itemize}

\subsection{A Variant of Informed-MIA via Privacy Backdoor}\label{sec:backdoor-game-APP}
When demonstrating the dataset-level risk with variants of MIAs, we aim to use a more informed MIA game by assuming a stronger (informed) adversary knowledge, thus the mitigation under such strong privacy game can be reducible to other weaker attacks in practice.
Thus, we follow previous works of privacy backdoor~\cite{liu2024precurious, wen2024privacy} by assuming the adversary has the capability to craft and release the pre-trained model.
The informed MIA game is shown in \Cref{alg:informed-adv-backdoor-APP}.

\begin{algorithm}[H]
\caption{\textsc{Backdoored and Informed MIA Game}}
\label{alg:informed-adv-backdoor-APP}
\begin{algorithmic}[1]
\Procedure{\textsc{Backdoored-MIA}}{$\mathcal{T}, \mathcal{A}, D_{\backslash 
x}, \x; D_\text{aux}$}
    \State $\theta_\text{adv} \gets \mathcal{A}_{\text{craft}}(D_\text{aux}, \theta_\text{pre})$ \textcolor{gray}{\textit{// Insert privacy backdoor}}
    \State $s\gets \text{Unif}(\{\x, \bot\})$ \textcolor{gray}{\textit{// Sample membership status}}
    \State $\theta \gets \mathcal{T}(D_{\backslash \x} \cup \{\tilde{\x}|s\neq \empty \bot\} ; \theta_\text{adv})$ \textcolor{gray}{\textit{// Train on released data with the backdoored pre-trained model}}
    \State $\hat{s} \gets \mathcal{A}(\theta_\text{ft}, \theta_\text{adv}, \x; D_{\text{aux}})$ \textcolor{gray}{\textit{// Guess membership of $\x$}}
    \State \Return $\hat{s}=s$
\EndProcedure
\end{algorithmic}
\end{algorithm}

\section{Additional Results}

\subsection{Instance Exploitation}\label{APP:sec:exploitation}

\noindent\textbf{Additional Results for CC-News and Enron.}
\begin{figure}[thb]
    \centering
    
    \includegraphics[width=0.43\linewidth]{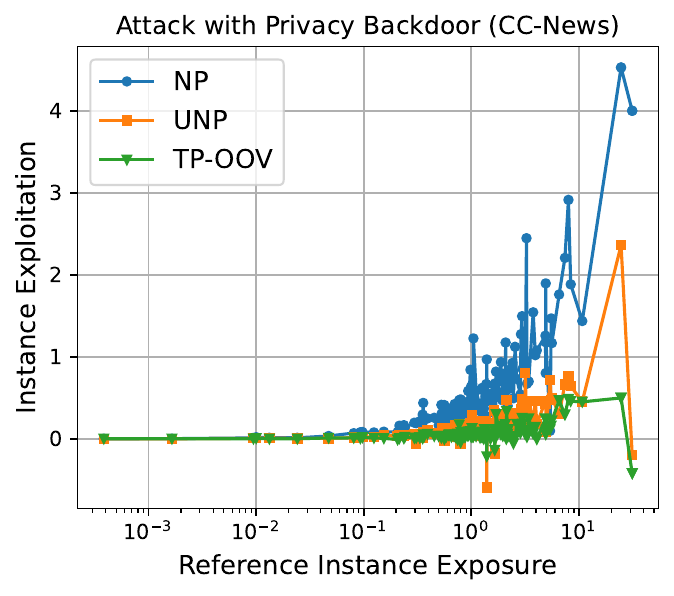}
    \includegraphics[width=0.45\linewidth]{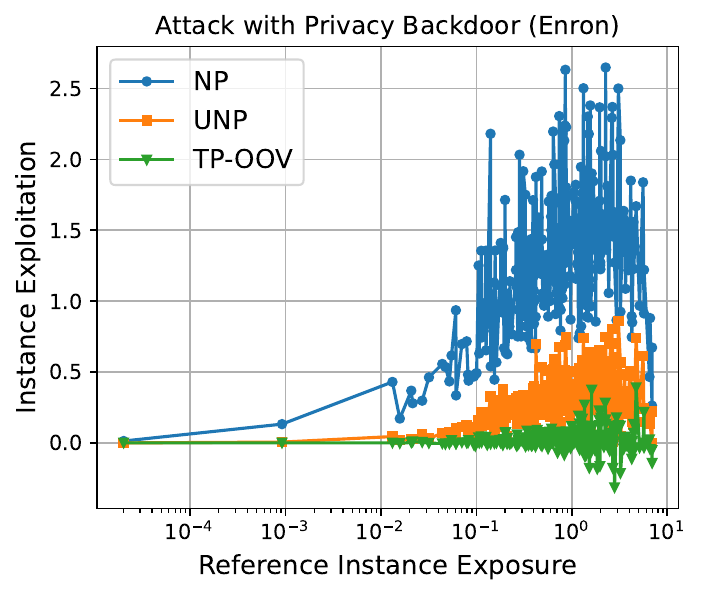}
    \caption{Instance-level analysis via instance exploitation, with the corresponding maximum MIA AUC as 
    0.605 for CC-News ($b=1$), and 0.621 for Enron ($b=1$) obtained with privacy backdoor.
    }
    \label{fig:instance-APP}
\end{figure}
We complement results of \Cref{fig:instance} on two extra datasets CC-News and Enron in \Cref{fig:instance-APP} with the similar trend that after applying \ours the instance exploitation for most text instances approaches to zero.
The gain of perturbing with OOV compared to uniform token sequence is significant for all datasets.

\noindent\textbf{Results with Different Budgets.}
\begin{figure}[thb]
    \centering
    \includegraphics[width=0.65\linewidth]{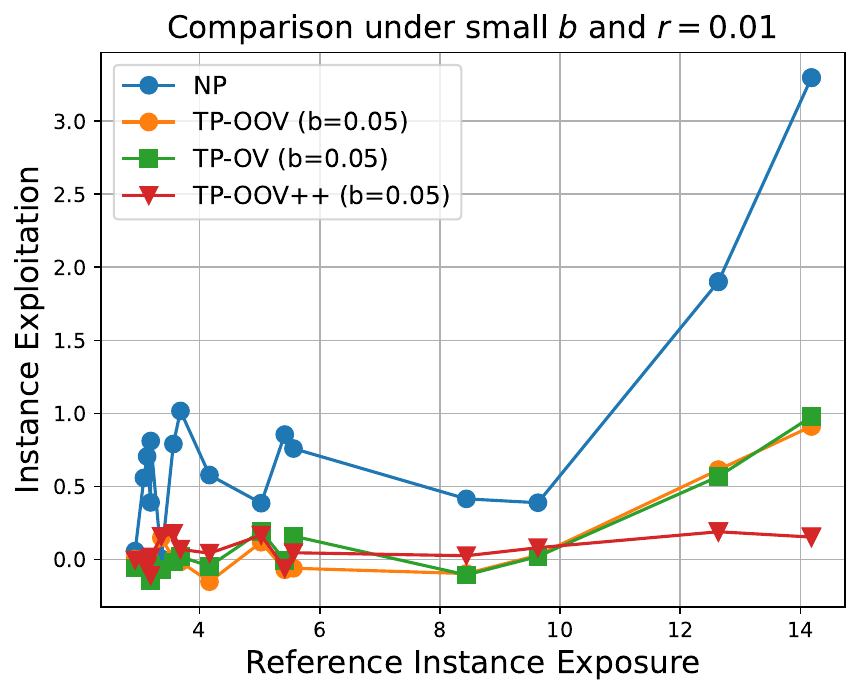}
    \caption{Effectiveness of optimization-based method on most vulnerable instances given a small portion of defender $r=0.01$ and small portion of perturbation budget.}
    \label{fig:eff_opt-APP}
\end{figure}
We complement results of \Cref{fig:eff_opt} with different perturbation budget $b$ in \Cref{fig:eff_opt-APP}.

\subsection{Perturbation for Data Watermark}\label{sec:watermark-APP} 
Although we target for proactive memorization mitigation while data watermark is a reactive strategy for data proving, we have similar technique of perturbing text in general.
Technically, \ours can be extended as data watermark for claiming the ownership.
Specifically, prior works~\cite{wei2024proving, zhaoprotecting} insert random canary into protected content, query the model with the canary, and then perform MIA for watermark detection.

Although watermarking is not our focus, we demonstrate \ours's feasibility as data watermark in \Cref{tab:effect_detect_APP}.
Using p-value and z-score metrics for hypothesis testing (lower values indicate stronger detection), \ours provides extremely strong detection power.
TP-P with artificial memorization tokens performs best among variants.
While TP-OP should theoretically excel, optimization across multiple samples proves time-consuming, yielding insufficient pitfall optimization.
Nevertheless, all variants detect data watermarks effectively.

\begin{table}[thb]
\centering
\caption{Detection effectiveness for Patient dataset and GPT-2 Model when \ours serves as data watermark.}
\label{tab:effect_detect_APP}
\resizebox{0.4\textwidth}{!}{
\begin{tabular}{c|cc|cc}
\toprule[1pt]
          & \multicolumn{2}{c}{Detection w/o context} & \multicolumn{2}{c}{Detection w/ context} \\
Method    & p-value              & z-score            & p-value             & z-score            \\
\midrule[0.8pt]
UDP     & 6.90E-07             & -12.858            & 3.57E-07            & -12.122            \\
UNP    &       2.34E-17 & 	-20.696 	 & 2.29E-22 &	-25.265        \\
TP     & 2.74E-11             & -19.769            & 1.75E-292           & -122.483           \\
TP-P & 1.90E-87             & -48.648            & 4.14E-305           & -101.609          \\
TP-OP & 7.47E-09             & -15.860            & 1.29E-233           & -116.993           \\
\bottomrule[1pt]
\end{tabular}
}
\end{table}

\subsection{MIA Results}
\noindent\textbf{Influence of Defender Portion.}
We demonstrate the influence of defender portion $r$ on the defense effectiveness in \Cref{fig:inf_r_APP}.
A smaller defender ratio results in a lower privacy risk, or equally a better defense effectiveness under both sample-level and user-level MIA.

\begin{figure}
\centering
    \includegraphics[trim=20 0 0 0, clip, width=0.3\textwidth]{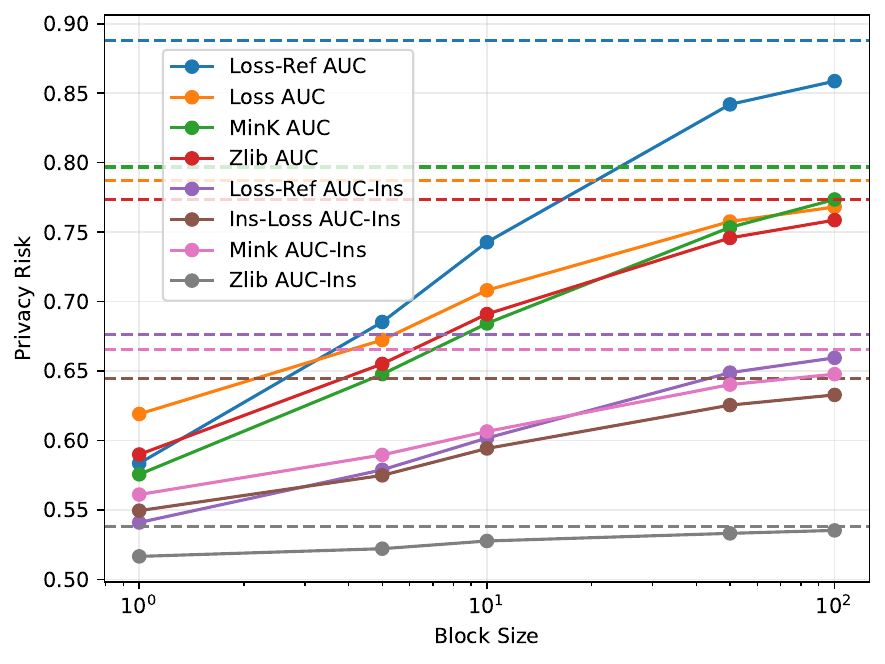}
    \caption{Influence of disturbing in UDP.}
	\label{fig:inf_freq}
\end{figure}

\noindent \textbf{Full Results of Variant MIA Signals.}
Due to space limitation, we omit the MIA results for each MIA signal and only report the maximum MIA AUC and TPR across different MIA signals.
We report the full results with privacy backdoor \Cref{tab:mainbd-APP} and omit the one without privacy backdoor due to space limitation.

\begin{table}[thb]
\centering
\caption{Influence of defender portion $r$ for UDP.}
\label{fig:inf_r_APP}
\resizebox{0.4\textwidth}{!}{
\begin{tabular}{c|cc|cc}
\toprule
UDP                & \multicolumn{2}{c}{Sample-Level} & \multicolumn{2}{c}{User-Level} \\
\midrule
Defender Portion r & AUC            & TPR@1\%         & AUC           & TPR@1\%        \\
\hline
1.000              & 0.791          & 0.068           & 0.616         & 0.008          \\
0.800              & 0.775          & 0.112           & 0.612         & 0.011          \\
0.500              & 0.779          & 0.130           & 0.598         & 0.009          \\
0.100              & 0.743          & 0.091           & 0.602         & 0.000          \\
0.050              & 0.734          & 0.059           & 0.601         & 0.078         \\
\bottomrule
\end{tabular}
}
\end{table}

\begin{figure}[tbh]
\centering
    \includegraphics[width=0.3\textwidth]{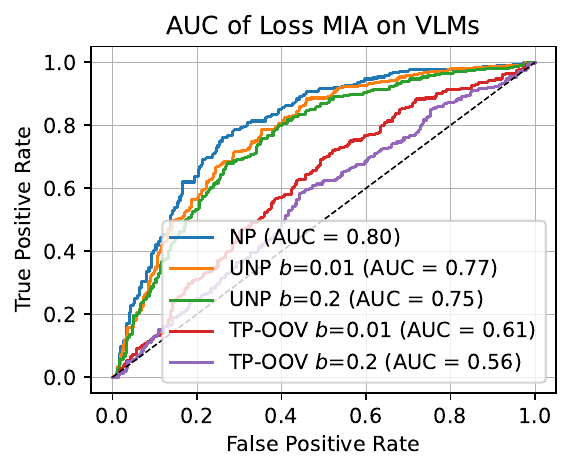}
    \caption{Loss MIA performance on VLMs.}
	\label{fig:vlm-loss-mia}
\end{figure}

\begin{table*}[t]
\centering
\caption{Membership inference evaluation with privacy backdoor.}
\label{tab:mainbd-APP}
\resizebox{0.8\textwidth}{!}{
\begin{tabular}{c|c|cc|cc|cc|cc}
\toprule
\multirow{1}{*}{MIA level} & Patient GPT-2   & \multicolumn{2}{c}{Loss-Reference} & \multicolumn{2}{c}{Loss} & \multicolumn{2}{c}{MinK} & \multicolumn{2}{c}{Zlib} \\
\midrule
                      w/ Backdoor                 & Method          & AUC           & TPR@1\%FPR         & AUC      & TPR@1\%FPR    & AUC      & TPR@1\%FPR    & AUC      & TPR@1\%FPR    \\
\midrule
\multirow{7}{*}{Sample-level}          & NP              & 0.953         & 0.545              & 0.792    & 0.364         & 0.815    & 0.303         & 0.783    & 0.303         \\
                                       & UDP (b=0.4)     & 0.831         & 0.121              & 0.710    & 0.152         & 0.704    & 0.121         & 0.697    & 0.182         \\
                                       & UNP (b=0.4)     & 0.766         & 0.091              & 0.674    & 0.152         & 0.679    & 0.091         & 0.658    & 0.152         \\
                                       & TP (b=0.4)      & 0.765         & 0.091              & 0.671    & 0.152         & 0.644    & 0.061         & 0.657    & 0.182         \\
                                       & TP-P (b=0.4)    & 0.772         & 0.091              & 0.676    & 0.152         & 0.654    & 0.061         & 0.655    & 0.182         \\
                                       & TP-OOV (b=0.4)  & 0.566         & 0.061              & 0.587    & 0.061         & 0.503    & 0.030         & 0.562    & 0.091         \\
                                       & UNP-OOV (b=0.4) & 0.648         & 0.091              & 0.648    & 0.121         & 0.614    & 0.061         & 0.624    & 0.152         \\
\midrule
\multirow{7}{*}{User-level}            & NP              & 0.741         & 0.000              & 0.652    & 0.047         & 0.672    & 0.047         & 0.540    & 0.023         \\
                                       & UDP (b=0.4)     & 0.649         & 0.000              & 0.599    & 0.047         & 0.611    & 0.039         & 0.528    & 0.023         \\
                                       & UNP (b=0.4)     & 0.622         & 0.000              & 0.582    & 0.039         & 0.598    & 0.039         & 0.524    & 0.023         \\
                                       & TP (b=0.4)      & 0.618         & 0.000              & 0.581    & 0.039         & 0.585    & 0.039         & 0.523    & 0.023         \\
                                       & TP-P (b=0.4)    & 0.619         & 0.000              & 0.580    & 0.023         & 0.592    & 0.039         & 0.523    & 0.023         \\
                                       & TP-OOV (b=0.4)  & 0.542         & 0.000              & 0.538    & 0.039         & 0.521    & 0.039         & 0.515    & 0.023         \\
                                       & UNP-OOV (b=0.4) & 0.566         & 0.000              & 0.566    & 0.039         & 0.574    & 0.039         & 0.520    & 0.023         \\
\toprule
\multirow{1}{*}{MIA level} & CC-News OPT-125M   & \multicolumn{2}{c}{Loss-Reference} & \multicolumn{2}{c}{Loss} & \multicolumn{2}{c}{MinK} & \multicolumn{2}{c}{Zlib} \\
\midrule
                    w/ Backdoor                   & Method          & AUC           & TPR@1\%FPR         & AUC      & TPR@1\%FPR    & AUC      & TPR@1\%FPR    & AUC      & TPR@1\%FPR    \\
\midrule
\multirow{7}{*}{Sample-level}          & NP              & 0.998         & 0.982              & 0.642    & 0.006         & 0.668    & 0.006         & 0.700    & 0.036         \\
                                       & UDP (b=0.4)     & 0.997         & 0.970              & 0.594    & 0.006         & 0.605    & 0.006         & 0.650    & 0.030         \\
                                       & UNP (b=0.4)     & 0.983         & 0.467              & 0.556    & 0.006         & 0.559    & 0.006         & 0.613    & 0.030         \\
                                       & TP (b=0.4)      & 0.978         & 0.580              & 0.554    & 0.006         & 0.548    & 0.006         & 0.610    & 0.024         \\
                                       & TP-P (b=0.4)    & 0.991         & 0.746              & 0.560    & 0.006         & 0.555    & 0.006         & 0.615    & 0.024         \\
                                       & TP-OOV (b=0.4)  & 0.890         & 0.083              & 0.518    & 0.006         & 0.510    & 0.006         & 0.577    & 0.024         \\
                                       & TP-OOV (b=1)    & 0.621         & 0.053              & 0.498    & 0.006         & 0.495    & 0.006         & 0.554    & 0.018         \\
\midrule
\multirow{7}{*}{User-level}            & NP              & 0.966         & 0.032              & 0.620    & 0.032         & 0.648    & 0.035         & 0.551    & 0.007         \\
                                       & UDP (b=0.4)     & 0.948         & 0.028              & 0.582    & 0.032         & 0.607    & 0.035         & 0.542    & 0.007         \\
                                       & UNP (b=0.4)     & 0.912         & 0.028              & 0.559    & 0.032         & 0.581    & 0.032         & 0.535    & 0.007         \\
                                       & TP (b=0.4)      & 0.918         & 0.028              & 0.557    & 0.032         & 0.573    & 0.035         & 0.536    & 0.007         \\
                                       & TP-P (b=0.4)    & 0.923         & 0.032              & 0.559    & 0.032         & 0.577    & 0.035         & 0.536    & 0.007         \\
                                       & TP-OOV (b=0.4)  & 0.783         & 0.028              & 0.532    & 0.032         & 0.547    & 0.035         & 0.531    & 0.007         \\
                                       & TP-OOV (b=1)    & 0.605         & 0.028              & 0.521    & 0.032         & 0.543    & 0.035         & 0.527    & 0.007        \\
\bottomrule
\end{tabular}
}
\end{table*}

\subsection{Vision-Language Modelling}
\label{app:vlm}
\noindent\textbf{Setup.} We use IAPR TC-12 dataset~\cite{iapr} covering diverse subjects (sports, people, animals, cities, landscapes) with image-caption pairs including title, description, location, and date. We focus on image captioning using BLIP2-ViT-gOPT2.7B~\cite{blip2} (3.8B parameters) with standard causal language modeling. 
The setup takes an image and partially masked caption as input, outputting the next caption word. 
We employ LoRA (rank=16) on query and value matrices across vision encoder, Q-former, and LLM components for efficiency. We use 3K image-caption pairs for $D$ (10\% protected as $D_\text{pro}$), train reference models on separate 3K pairs, and optimize for 20 epochs using AdamW (lr=5e-4) with validation on 500 images.

\noindent\textbf{Results of Loss MIAs.} TP-OOV effectively reduces MIA performance, consistent with main paper results. 
\Cref{fig:vlm-loss-mia} shows ROC curves where baseline NP marginally affects MIAs (AUC: 0.8→0.77/0.75 for $b=0.01/0.02$), while TP-OOV significantly reduces AUC to 0.56 ($b=0.2$) and 0.61 ($b=0.01$) from 0.80.

\subsection{Ethical Considerations}
This work presents a proactive defense to safeguard users' released text from potential LLM misuse, aligning with the ethical principles of "Respect for Persons" and "Beneficence" (e.g., as outlined in the Menlo Report~\cite{dittrich2012menlo}) by promoting individual autonomy and data privacy.

While providing a valuable safeguard, our text modification technique introduces dual ethical considerations. First, the mechanism could be maliciously exploited by adversaries to embed imperceptible modifications, potentially leading to data poisoning, backdoor vulnerabilities, or performance degradation in trained models. 
Second, the pursuit of individual protection may inadvertently impact collective fairness. For instance, a data owner's successful defense marginally increases the likelihood of other users' unprotected text being exposed through model outputs. 

Therefore, we perform analysis related to above considerations.
1) Our extensive experiments show that training on a small portion of protected text does not degrade model performance. 
2) We discussed the collaborative mitigation in \Cref{sec:discussion} by removing the whole perturbed and protected content from training corpus. 
3) We demonstrated that the fairness issue is not observed given a reasonably small portion of protection set (with $r<0.5$ in \Cref{fig:inf_r_un}).
Meanwhile, we encourage users and practitioners must remain vigilant to these broader ramifications, ensuring system integrity and collective fairness are not inadvertently compromised.

\end{document}